\documentclass[fleqn,usenatbib]{mnras}
\usepackage{newtxtext,newtxmath}
\usepackage[T1]{fontenc}

\DeclareRobustCommand{\VAN}[3]{#2}
\let\VANthebibliography\thebibliography
\def\thebibliography{\DeclareRobustCommand{\VAN}[3]{##3}\VANthebibliography}

\usepackage{graphicx}	% Including figure files
\usepackage{amsmath}	% Advanced maths commands

\usepackage{bm}		% Bold maths symbols, including 
\usepackage{colortbl}
\usepackage{xcolor}
\definecolor{sof}{rgb}{0.4, 0.0, 0.6}

\definecolor{mgm}{rgb}{0.0, 0.0, 1.0}

\title[Nulling Pulsar Statistics]{A Statistical Analysis of the Nulling Pulsar Population}

\author[Sheikh and MacDonald]{
Sofia Z. Sheikh,$^{1, 2}$\thanks{E-mail: szs714@psu.edu (SZS)}
Mariah G. MacDonald,$^{1, 2}$
\\
$^{1}$Department of Astronomy \& Astrophysics, The Pennsylvania State University, University Park, PA 16802, USA\\
$^{2}$Center for Exoplanets and Habitable Worlds, The Pennsylvania State University, University Park, PA 16802, USA\\
}

\date{Accepted XXX. Received YYY; in original form ZZZ}

\pubyear{2020}

\begin{document}
\label{firstpage}
\pagerange{\pageref{firstpage}--\pageref{lastpage}}
\maketitle

\begin{abstract}
Approximately 8\% of the $\sim$2800 known pulsars exhibit ``nulling,'' a temporary broadband cessation of normal pulsar emission. Nulling behaviour can be coarsely quantified by the \textit{nulling fraction}, which describes the percentage of time a given pulsar will be found in a null state. In this paper, we perform the most thorough statistical analysis thus far of the properties of 141 known nulling pulsars. We find weak, non-linear correlations between nulling fraction and pulse width, as well as nulling fraction and spin period which could be attributed to selection effects. We also further investigate a recently-hypothesized gap at 40\% nulling fraction. While a local minimum does exist in the distribution, we cannot confirm a consistent and unique break in the distribution when we investigate with univariate and multivariate clustering methods, nor can we prove the existence of two statistically distinct populations about this minimum. Using the same methods, we find that nulling pulsars are a statistically different population from normal, radio, non-nulling pulsars, which has never been quantitatively verified. In addition, we summarize the findings of the prior nulling pulsar statistics literature, which are notoriously contradictory. This study, in context, furthers the idea that nulling fraction alone does not contain enough information to describe the behaviour of a nulling pulsar and that other parameters such as null lengths and null randomness, in addition to a better understanding of selection effects, are required to fully understand this phenomenon. 
\end{abstract}


%%%%%%%%%%%%%%%%%%%%%%%%%%%%%%%%%%%%%%%%%%%%%%%%%%

%%%%%%%%%%%%%%%%% BODY OF PAPER %%%%%%%%%%%%%%%%%%

\section{Introduction}

Pulsar ``nulling'' \citep[first reported in][]{backer1970pulsar} is a relatively-common phenomenon where otherwise strong and steady radiation from a pulsar decreases by at least a factor of 100,\footnote{Factors up to 1420 have been found \citep[e.g., B1706-16, ][]{naidu2018detection}} and possibly entirely, over all frequencies \citep{ritchings1976pulsar, wang2007pulsar}. These nulls occur for durations of a single pulse period all the way up to days, and affect all components of a pulse simultaneously, (e.g. both core and conal components \citep{Rankin1986}, interpulses \citep{hankins1984psr1944+}). 

Nulling pulsars can be identified by looking for gaps in emission in phase-time diagrams \citep{wang2007pulsar}. The nulling fractions (NFs) can then be measured with pulse energy distributions, which will have an overabundance of pulses near zero energy, sometimes leading to a separate well-resolved peak in pulsars with large NF. These overabundances can be measured and reported as a percentage of the total number of pulses that are nulled \citep{ritchings1976pulsar}.

There are some caveats to this method. Firstly, nulling fraction is only one of the parameters that characterizes nulling behaviour. Other parameters, including the duration of a nulling episode and distribution of these durations, are needed to fully describe the nulling behaviour of a pulsar \citep{gajjar2012survey}. Secondly, sensitivity limits make it difficult to identify whether pulsars are entirely dormant during a null, or whether there is some residual low-level emission \citep{wang2007pulsar}. 

Some nulling pulsars are reported to show changes in their energetics before, after, or during a pulse \citep[e.g. an increase in pulse energy after a null or a slow-down rate that drops during a null,][]{backer1970pulsar,kramer2006periodically}, suggesting that nulls, like other phenomena such as mode-changing\footnote{The likely instability-induced switching from one primary pulse profile to other secondary ones \citep{wang2007pulsar}}, occur due to changes in a pulsar's entire magnetospheric environment \citep{Lyne1982, Lyne2010, GajjarJoshi2014}. This, when considered with the simultaneous broadband cessation during a null, implies a non-uniform flow of particles into the emission region or a global breakdown of the emission mechanism \citep[in a ``tipping point'' like instability, ][]{wang2007pulsar,Gajjar2014}, and not changes in relative geometry.

Determining correlations between nulls and other pulsar properties and behaviour allows for characterization and investigations of the pulsar emission mechanism and its evolution over time. Pulsar characteristic age, pulse morphology, and pulse periods have been closely investigated in relation to NF, but there is currently no consensus about their impact on NF. Some studies have reported a positive correlation between characteristic age and nulling fraction \citep{ritchings1976pulsar, wang2007pulsar}, interpreting this as a sign that pulsars null more and more as they age until they ``die'' and stop emitting entirely \citep{ritchings1976pulsar}. Other studies find no correlation with age at all \citep{biggs1992analysis}, or no correlation with age within a morphological group \citep{Rankin1986}. Using classes from \citet{Rankin1983}, \citet{Rankin1986} suggested that different pulse morphologies lead to different NFs (with triple and multiple profiles having the highest NFs). However, even using the same classes, \citet{wang2007pulsar} did not find such a correlation and suggested that the known relationship between complex pulse morphology and older age \citep{huguenin1971properties} could have caused the previously observed correlated behaviour. Suggestions of a correlation between NF and period \citep{ritchings1976pulsar, biggs1992analysis} are hard to entangle from the previous discussion of correlations with age and morphology, and are also debated in the literature \citep{Rankin1986}.

The statistical methods used in the majority of this previous literature have been preliminary at best. Most studies graphed relevant variables against each other and looked for trends via visual inspection, though they were working with much less data and often admitted the limitations to claiming a correlation. \citet{Rankin1986} calculated means and standard deviations of different pulsar populations to claim that they were distinct. Some studies, specifically \citet{biggs1992analysis} and \citet{Li1995}, attempted to account for the upper limits and uncertainties on the nulling fraction measurements by using survival analysis methods such as Cox's proportion hazard model \citep{cox1972regression} and Kendall's rank correlations \citep{kendall1938new}. Most recently, \citet{Konar2019} quantified correlations with nulling fraction using Pearson's R.

\citet{Konar2019} provided the results of a thorough literature survey for NF values and the most recent statistical survey of nulling pulsar properties. The authors tentatively propose a gap in NF values around 40\% and claim that a Kolmogorov-Smirnov test on the spin-period of nulling pulsars with NF values greater than and less than 40\% imply that there are two distinct populations of pulsars \citep{Konar2019}. The authors also did not find correlations between NF and any intrinsic pulsar parameters, in contrast with previous studies.

In this paper, we will expand upon the table of pulsar parameters provided by \citet{Konar2019}, investigate correlations between intrinsic pulsar parameters and null fractions, compare the nulling and non-nulling pulsar populations, and statistically characterize and attempt to confirm the two populations around the suggested 40\% NF gap from \citet{Konar2019}. In Section~\ref{sec:data}, we explain how we compile and prepare our data and define all of the parameters that will be used in the statistical analysis described in Section~\ref{sec:methods}. We then explore the existence of the two distinct populations in null fraction suggested by \citet{Konar2019} in Section~\ref{sec: populations}, and look at trends between nulling fraction and morphological class in Section~\ref{sec: morphology}.  We provide the results from the correlation analyses between the nulling fraction and between other parameters in the nulling pulsar population in Section \ref{sec: correlations}. In Section \ref{sec: nullvsnonnull}, we look at the differences between the nulling and non-nulling pulsar populations before discussing our results in Section~\ref{sec:discussion} and concluding in Section~\ref{sec:conclusion}.
\newpage

\section{Data} 
\label{sec:data}

The data used in the analyses in Sections \ref{sec: populations} -- \ref{sec: nullvsnonnull} are comprised of 15 pulsar parameters and derived quantities. We show 14 of the 15 parameters used in the study in Table \ref{tab:data_table}. We discuss the remaining parameter, morphology, below. We gathered these parameters for a set of 141 pulsars contained in Tables 1--5 of \citet{Konar2019}, excluding pulsars in their sample without a measured nulling fraction. Pulsars with 2 or 3 measurements of nulling fraction in the literature have 2 or 3 rows in the table, respectively, for a total of 162 rows\footnote{There are actually 163 individual estimates of nulling fractions, but the estimate for B1713-40 from \citet{Wang2007} was accidentally omitted in our analyses.}. In our analyses, we use all 162 measurements of nulling fraction, as if each measurement were from a separate pulsar. Table \ref{tab:data_table} includes a description of each parameter and an explanation of how it was collected.

In addition to the parameters shown in Table \ref{tab:data_table}, we compile \citet{Rankin1983} pulse morphology classes for 141 of the 162 pulsar entries in our sample. 120 of these have been previously classified, and we assigned classifications to 22 additional pulsar entries based on published pulse profiles and analyses. These classifications are: S$_d$ (conal-component single-peaked profile), S$_t$ (core-component single-peaked profile), D (double-peaked profile), T (triple-peaked profile), and M (multiple profile containing more than three peaks in structure). We added an additional letter U (``undetermined'') for the remaining 20 pulsars and did not consider them in the morphology analysis in Section~\ref{sec: morphology}.

\begin{table*} 
    \centering
    \begin{tabular}{p{2.75cm}p{1cm}p{1cm}p{2cm}p{3cm}p{3cm}}
        \hline
        Name & Symbol & Units & Equation & Data Reference & Description  \\
        \hline
        
        Pulse Period & $P$ & s & N/A & \citet{Konar2019}, from \citet{Manchester_2005} & Barycentric spin period \\
        
        Spindown Rate & $\Dot{P}$ & s/s & N/A & \citet{Manchester_2005} & First time derivative of pulse period \\
        
        Magnetic Flux & $B_s$ & G & $B_s \propto (P \Dot{P})^{1/2}$ & \citet{Konar2019}, from \citet{Manchester_2005}, \citet{ostriker1969nature}* & Surface magnetic flux density \\
        
        Dispersion Measure & DM & cm$^3/$pc & N/A & \citet{Manchester_2005} & Indirect measure of pulsar distance from frequency-dependent arrival time delay \citep[e.g., ][]{jokipii1969faraday} \\
        
        Characteristic Age & $\tau_c$ & yr & $\tau_c \sim \frac{P}{2\Dot{P}}$ & Existing table values & Measure of pulsar age from period and spindown rate \\
        
        Kinetic Age & $\tau_k$ & yr & $\tau_k = \frac{z}{v_{z}}$ & Derived with $z$ and $v_{z}$ from \citet{Manchester_2005}, using \texttt{AstroPy} \citep{robitaille2013astropy} & Period-independent pulsar age derived from pulsar's distance from the galactic plane and z-velocity away from the plane$^\dagger$\\
        
        Radio Luminosity & $L_{R400}$ & mJy/kpc$^2$ & $L_{R400} \propto S_{400} d^2$ & Derived with $d$ and $S_{400}$ from \citet{Manchester_2005} & Radio luminosity at 400 MHz, quantifies intrinsic pulsar power\\
        
        Spindown Luminosity & $\Dot{E}$ & mJy/kpc$^2$ & $\Dot{E} \propto \Dot{P}P^{-3}$ & \citet{Manchester_2005}, \citet{gold1969rotating}* & Spindown energy loss rate\\
        
        Maximum Magnetic Flux Density & $B_{lc}$ & G & $B_{lc} \propto P\Dot{P}^{-5}$ & Existing table values, \citet{goldreich1969physics, lyne1975period}* & Magnetic field in the ``light cylinder'' region of the pulsar magnetosphere, where particle velocities are near c\\
        
        Plasma Flow Parameter & $Q$ & None & $Q \propto P^{1.1}\Dot{P}^{-0.4}$& Existing table values, \citet{beskin1984spin}* & Derived parameter empirically related to pulsar morphology, with $Q<<1$ generally indicating a single-pulse morphology \citep{beskin2009mhd}\\
        
        Pulse Width & $W$ & ms & N/A & \citet{Manchester_2005} & Width of the average pulse profile measured at 50\% and 10\% of the peak brightness\\
        
        Magnetic Inclination Angle & $\alpha$ & $\circ$ & N/A & \citet{Malov1990, malov2011angles, Nikitina2017}$^\ddag$ & Angle between spin axis of the pulsar and magnetic moment\\
        
        Nulling Fraction & NF & \% & N/A & Tables 1--5 of \citet{Konar2019} & Percentage of time that a pulsar is in a nulled state\\
        \hline
    \end{tabular}
    \caption{14 of the 15 pulsar parameters used in the following statistical analyses. Equations are only given for derived parameters. References marked with * give the equations used for derived parameters. \newline \newline
    $^\dagger$ These data were only available for 96 pulsars in the sample. It should be noted that pulse widths change significantly when measured at different frequencies, or with different time resolutions, and that any use of these data must take into account the large additional uncertainties introduced to a potential correlation. \newline \newline
    $^\ddag$We compiled $\alpha$ values (reported as $\beta$ in some references) from the literature for 107 pulsars in our sample. As noted by \citet{Nikitina2017}, $\alpha$ can be calculated from pulse widths \citep[assuming that the line of sight passes through the center of the radiation cone as in][]{rankin1990toward}, polarization information from the main cone, or position angle and mean profile shape, as in \citet{lyne1988shape}. The reported $\alpha$ values used here are averages from the different methods, where possible. It can be difficult to measure $\alpha$ values with the pulse width method for pulsars with extremely narrow pulse widths \citep[e.g., J1717-4054, ][]{Kerr2014}.}
    \label{tab:data_table}
\end{table*}

\section{Methods}\label{sec:methods}

\subsection{Cluster Analyses}
\label{ssec: clustering}

Cluster analyses are a set of methods which can determine how many distinct subsets lie within a single dataset, which points belong to each of the subsets, and whether the subsets themselves are statistically distinct. We employ multiple different clustering techniques, described below.

\subsubsection{Kernel Density Estimation}
\label{sssec: kde}

A kernel density estimation (KDE) is a nonparametric estimation of probability density functions, defined by:

\begin{equation}
    \hat{f}(x;H) =n^{-1}\sum^n_{i=1}K_H(x-X_i)
\end{equation}
\noindent where $x=(x_1,x_2)^T$ and $X_i = (X_{i1},X_{i2})^T$ for $i$=1,2,...,n a random sample drawn from a density $f$, $K(x)$ is the kernel which is a symmetric probability density function, and $H$ is the positive-definite and symmetric bandwidth matrix. We use an axis-aligned normal kernel. When estimating the location of the gap in nulling fraction, we use a variety of bandwidths from  U[1.0,100.0] (see Section~\ref{sec: populations} for full analysis).

\subsubsection{Jenks' Natural Breaks Optimization}
\label{sssec: jenks}
Jenks' Natural Breaks Optimization (JNBO) is a univariate class interval method originally developed for the classifying geographic data for cartography \citep{jenks1977optimal}. Now it is used in many fields beyond geography to find a user-specified number of homogeneous classification groups in one-dimensional data \cite[e.g., ][]{rahadianto2015risk}. Once the number of classes has been defined, the JNBO algorithm tries to iteratively minimize the sum of the variances from the class means \citep{coulson1987matter}. Jenks-Fisher is a more runtime-efficient version of this same algorithm, incorporating the method put forth in \citet{fisher1958grouping} \citep{classInt2019}.

\subsubsection{Hierarchical Clustering}
\label{sssec: hierarchical}

Hierarchical clustering is a clustering method which does not immediately produce a user-specified number of clusters. Agglomerative algorithms such as \texttt{hclust} in R instead look for the two ``closest'' points (by some linkage metric) and then combine them into a cluster, iteratively repeating until all points are incorporated into a single cluster \citep{leisch1999bagged, mullner2013fastcluster}. The results for any number of clusters are thus pre-calculated. The main drawback of hierarchical clustering is its large computational cost, but its ability to use different linkage metrics makes it a particularly flexible solution for clustering \citep{leisch1999bagged}.

\subsubsection{Bagged Clustering}
\label{sssec: bagged}

Bagged (``\textbf{b}ootstrap \textbf{agg}regat\textbf{ed}'') clustering is a clustering method that is both a partitioning method, like k-means clustering \citep{macqueen1967some}, and a hierarchical method (as described in Section \ref{sssec: hierarchical}) \citep{leisch1999bagged}. Partitioning methods such as k-means are less effective at recovering structure if the clusters are not convex and, as mentioned in the previous section, hierarchical clustering can be computationally intensive; bagged clustering algorithms such as \texttt{bclust} in R avoid both drawbacks \citep{leisch1999bagged}.

\subsubsection{Geometrical Interval Classification}
Geometrical interval classification classifies data into intervals that have a geometric series. The algorithm forms these intervals by minimizing the sum of the squares of the number of points in each class. Interval classifications also aim for each class range to be approximately the same size and for the change between intervals to be consistent \citep{xu2002}. This algorithm is designed to accommodate continuous data and looks to highlight changes in the data at extreme values as well as at moderate values, making it ideal for heavily skewed and non-normal distributions.

\subsection{Kolmogorov-Smirnov and Anderson-Darling Tests}
\label{ssec: ks_ad}

The Kolmogorov-Smirnov (K-S) test is a nonparametric test of one-dimensional probability distributions that can be used to compare two samples. The K-S statistic quantifies the distance between the empirical distribution functions of the two samples and is 

\begin{equation}
    D_{n,m} = sup |F_{1,n}(x)-F_{2,m}(x)|
\end{equation}

\noindent where $F_{1,n}$ and $F_{2,m}$ are the empirical distribution functions of the two samples, and $sup$ is the supremum function. The null hypothesis that the two samples are from the same underlying population is rejected at a level $\alpha$ if

\begin{center}
$D_{n,m}> c(\alpha)\sqrt{\frac{n+m}{nm}}$, 

where

$c(\alpha) =\sqrt{-\frac{1}{2} \textrm{ln} \Big ( \frac{\alpha}{2} \Big )}$
\end{center}

Astronomers have been using the K-S test, for two-samples as well as for goodness-of-fit, for decades now \citep[e.g., ][]{peacock1983two}. This is mainly because it easy to compute and quite simple to understand, but the test is also distribution-free and can be universally applied without restrictions to sample size. Unfortunately, the K-S test is insensitive when the differences between the empirical distribution functions are not in the middle, but rather at the beginning and the end.

The Anderson-Darling (A-D) test oftentimes tests for normality, or compares a sample's empirical distribution function to that of another known form (e.g. log-normal, exponential). Using the A-D measure of agreement between distributions, though, it is possible to nonparametrically test a sample, or multiple samples, to see if they come from the same distribution.

The test is based on the distance between the two empirical distribution functions:

\begin{equation}
    A^2 = n \int^{\infty}_{-\infty} \frac{(F_n(x)-F(x))^2}{F(x)(1-F(x))}dF(x)
\end{equation}

Like the K-S test, the A-D test is both nonparametric and distribution free, but it does not suffer from the same insensitivity issues as the K-S test since it compares the two functions over the entire interval. Also like the K-S test, the null hypothesis states that the two tested samples are from the same underlying population. In comparing two populations, we employ both K-S ad A-D tests.

\subsection{Pearson's R and the Maximal Information Coefficient}
\label{ssec: r_mic}
The Pearson correlation coefficient, also known as ``Pearson's R'' or bivariate correlation, measures the linear correlation between two different variables \citep{pearson1895vii}. The coefficient ranges between -1 and 1, where 0 indicates no correlation, 0.3--0.5 indicates a weak correlation, 0.5--0.7 indicates a moderate correlation, and greater than 0.7 indicates a strong correlation; negative values of the same ranges indicate anti-correlations. The Pearson correlation coefficient is used widely across scientific fields when analyzing data. The coefficient is the covariance of the two variables being tested, divided by the product of their standard deviations. Since we often do not know the entire population and merely study samples of the population, the coefficient is of the sample and calculated by:

\begin{equation}
r=r_{xy}=\frac{\Sigma x_iy_i-n\bar{x}\bar{y}}{(n-1)s_xs_y}
\end{equation}

\noindent where $n$ is the number of data points in the sample, $x_i$ and $y_i$ are each data point, $\bar{x}$ and $\bar{y}$ are the sample means for x and y, and $s_x$ and $s_y$ are the sample standard deviations for x and y.

Although Pearson's correlation coefficient is widely used, the coefficient only measures how linear a correlation is; it is therefore sometimes insufficient since two variables can be highly correlated without their correlation being linear. Many correlations in astronomy are power laws or exponential, and Pearson's correlation coefficient is insensitive in these situations. As such, we also analyze the data by looking at the Maximal Information Criteria \citep[MIC, ][]{reshef2011detecting}. Similar to Pearson's correlation coefficient, the MIC ranges between 0 and 1, with 0 indicating no correlation, and higher values corresponding with stronger correlations. The MIC is defined as:

\begin{equation}
    \textrm{MIC(D)}  = \genfrac{}{}{0pt}{}{\textrm{max}}{XY<B(n)} \frac{I^*(D,X,Y)}{\textrm{log}(\textrm{min}(X,Y))}
\end{equation}
\noindent where $D=D(x,y)$ is the set of $n$ ordered pairs of elements of $x$ and $y$ in a space partitioned by an $X\times Y$ grid, grouping the $x$ and $y$ values in $X$ and $Y$ bins, respectively, $B(n) = n^{\alpha}$ is the search-grid size, with the $\alpha$ parameter defined by the sample size  \citep[here, $\alpha=0.75$ ][]{albanese2018practical}, and $I^*(D,X,Y)$ is the maximum mutual information over all grids $X\times Y$ of the distribution induced by $D$ on a grid having $X$ and $Y$ bins.  

The MIC and other maximal information-based nonparametric exploration methods were introduced by \citet{reshef2011detecting}, who showed that the MIC for a random relationship was approximately 0.18, with anything near or below that number indicating no correlation. For this work, we do not use an MIC cut-off, but instead rely on p-values generated as described in Section~\ref{ssec: null_correlations}.

\subsection{Managing Upper Limits and Uncertainties}
\label{ssec: upperlims_uncertainties}

NF is often reported in the literature with associated upper and lower limits. In addition, some pulsars have only been seen to null once, leading to a NF that is only a moving upper limit until the pulsar is observed to null again. These uncertainties and upper limits must be addressed to ensure that any discovered correlations are statistically significant, so we account for them in the following manner. For pulsars with an uncertainty and a nominal value, we draw NF values for each pulsar from a normal distribution centered on the nominal value, with a standard deviation of the published uncertainty. If the published NF is an upper limit, we instead draw the NF from a log uniform distribution of U[0,$NF_u$], where $NF_u$ is the upper limit of the NF. We chose the log uniform distribution to account for the skewness of the NF distribution. We test the uniform distribution as well to ensure that the choice of distribution is not significantly affecting the results, as discussed in Section \ref{ssec: null_correlations}.

We produce an uncertainty-adjusted value for each NF as described above. When exploring correlations between NF and other parameters, we produce 1000 of these sets of corrected NFs and calculate the median of the correlation measurements from each of these 1000 sets. To evaluate the significance of that median, we permute our data as explained in Section \ref{ssec: null_correlations}.

\section{Two Different Populations about a Nulling Fraction of $\sim$ 40\%}
\label{sec: populations}

\subsection{Forming two populations}

\begin{figure} 
    \centering
    \includegraphics[width = 0.47 \textwidth]{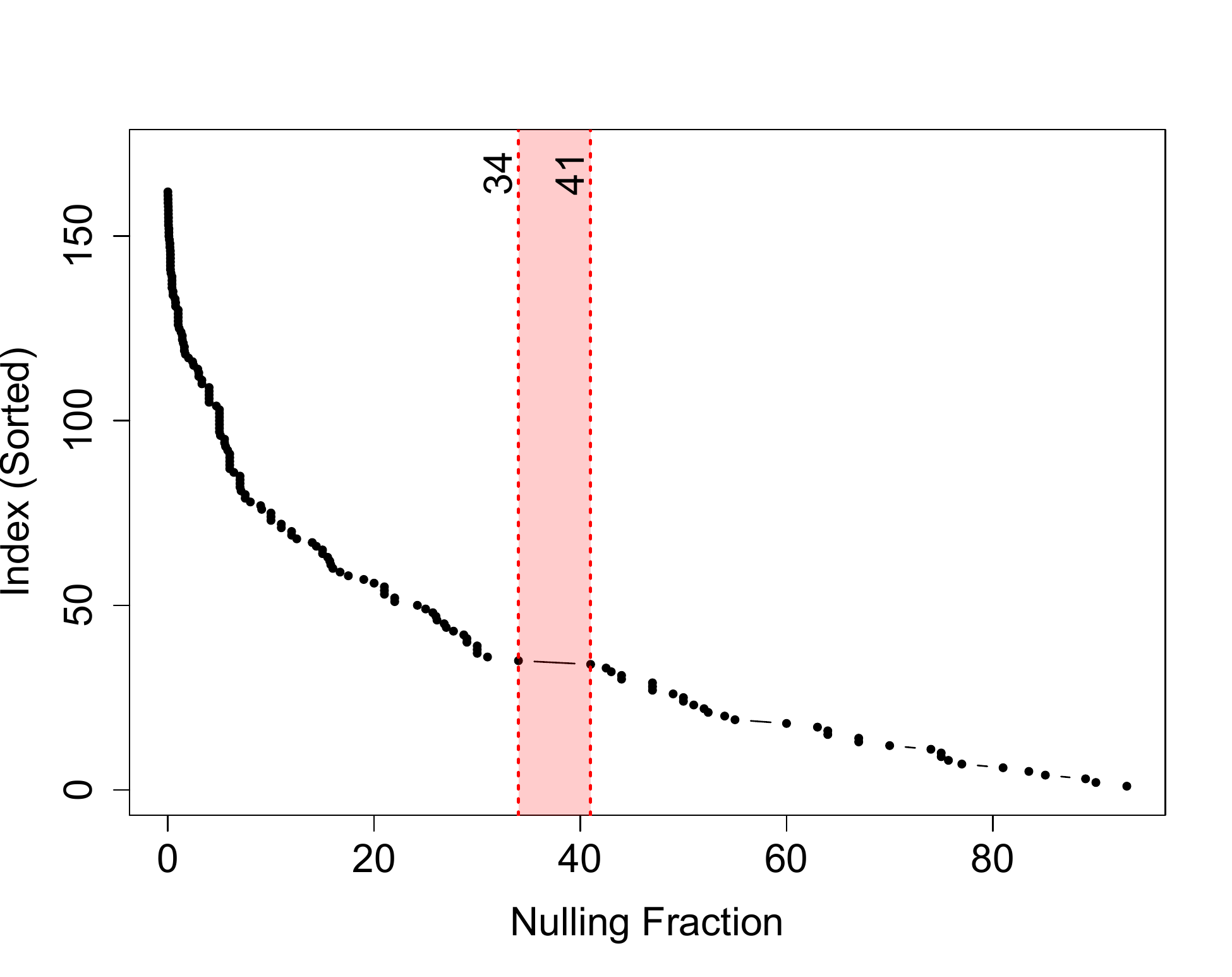}
    \caption{Results of Jenks Natural Breaks Optimization (Sections \ref{sssec: jenks}) with two classes run on the uncorrected nulling fraction distribution of 162 pulsar entries. All methods identified the class breakpoint to be in the gap between 30\% and 45\% for the uncorrected NF distribution, as well as for 1000 sets of corrected NF, in accordance with \citet{Konar2019}. Bagged clustering and hierarchical clustering also identified a secondary breakpoint at ~58\%. Geometrical interval classification, which does not assume a gap as the other methods do, concludes that both the corrected and uncorrected NF distributions are continuous and absent of true breaks. This suggests that the gap seen at ~40\% is an artifact of small sample size.}
    \label{fig:univariate_gap_final}
\end{figure}

Recently, \citet{Konar2019} suggested a gap in nulling fraction (NF) at $\sim40$\%. There are two separate claims that we would like to test: 1) that a ``gap'' or dearth of data exists at 40\% and 2) that this gap separates two statistically distinct populations. To confirm and characterize this gap, we use the CRAN package \texttt{classInt}\footnote{All statistics in this paper were performed using R 3.4.1 \citep{R2019}.} \citep{classInt2019} to run a series of univariate class interval algorithms with two classes on 1000 sets of corrected NF (see Section \ref{ssec: clustering} for a description of the methods). Hierarchical clustering, Jenks-Fisher, bagged clustering, and Jenks natural breaks opimization methods all identified the natural breakpoint in the data at $47.0\% \pm 11.2\%$, $34.4\% \pm 2.1\%$, $46.4\% \pm 12.4\%$, and $32.7\% \pm 1.5\%$, respectively. We show the result of a gap identification using Jenks natural breaks optimization in Figure~\ref{fig:univariate_gap_final}. Hierarchical clustering and bagged clustering alternate between identifying two different minima in the data: one at ~38\% and another at ~58\%, leading to the larger estimates and standard deviations. Geometric interval classification, using the CRAN package \texttt{cartography}, did not identify a break or gap near 40\%. In fact, the algorithm returned that all 1000 sets of corrected NF were continuous distributions, suggesting that the gap that we see and the breaks that are identified by the other methods are artifacts of an under-sampled population.

In order to directly compare to \citet{Konar2019}, we also test this method on the uncorrected NF, recovering gaps at ~37\% for each method, except geometric interval classification which again does not identify any breaks in the data.

We then estimate the distribution of NF using a 1D Kernel Density Estimator (KDE), using the uncorrected NF distribution. Using 100 bandwidths drawn from a uniform distribution U[1.0, 100.0]\footnote{Bandwidths drawn from U[1.0, 100.0] span the entire range of NF}, we confirm a minimum in the KDE of $37.23\pm0.19$. We draw multiple bandwidths such that the location of the minimum is not dependent on our bandwidth.

%We first examine this proposed gap using the uncorrected NF in order to compare to \citet{Konar2019}. However, as stated before, we must appropriately account for the uncertainties in the NF, including those reported as upper bounds. We repeat the series of class interval algorithms and KDE minimum identification with a corrected distribution of NF\footnote{Ideally, we would create a synthetic population of thousands of nulling pulsars with their NFs drawn from the known population, but class interval algorithms require a small number of points.} (see Section~\ref{ssec: upperlims_uncertainties}).
%With this new dataset, Jenks natural breaks optimization, Jenks-Fisher, and bagged clustering methods all identify the natural breakpoint in the data at 36\%, but hierarchical clustering is unable to identify any such gap. 

\begin{figure} 
    \centering
    \includegraphics[width=0.48\textwidth]{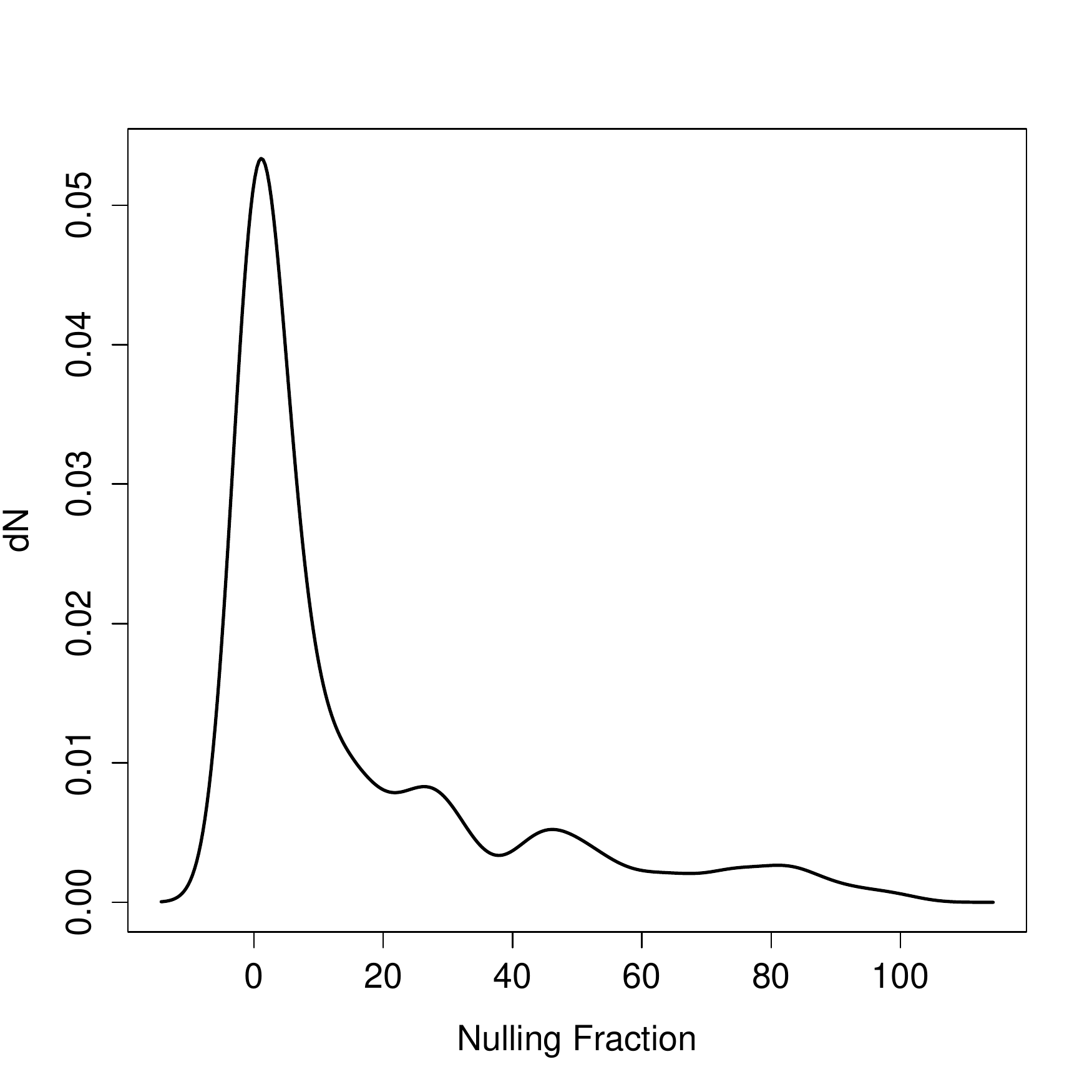}
    \caption{KDE of 16000 samples of corrected nulling fraction, with a bandwidth of 15.0. There does exist a local minimum in the distribution at around 40\%, as noted by \citet{Konar2019}, but this minimum is accompanied by two other local minima, at 20\% and at 60\%. This suggests that the gap seen in the data is due to small sample size.}
    \label{fig:kde}
\end{figure}

We then perform a KDE on 16000 samples of the corrected NF, again using 100 bandwidths drawn from a uniform distribution U[1.0, 100.0]. Regardless of bandwidth, there does exist a minimum in the KDE around 37\%, but there also exist minima at $\sim20$\% and at $\sim60$\% (see Figure~\ref{fig:kde}).
We thus find that the gap seen by \citet{Konar2019} is not a boundary between two distinct populations, but is simply an artifact of the uncertainties on NF estimates and an under-sampled population. Once uncertainties are accounted for, there does appear to be a local minimum at around 40\%, but other minima exist, suggesting that this minimum is solely due to the small sample size.

\subsection{Confirming two populations}
\label{ssec: confirming_two_pops}

We run both Kolmogorov-Smirnov and Anderson-Darling two-sample tests (see Section \ref{ssec: ks_ad}) on 1000 sets of corrected nulling fraction measurements. We define our two populations as pulsars with $NF<37.5\%$ and pulsars with $NF>37.5\%$.We compare the two populations in 13 parameters (P, $\dot{P}$, $B_s$, DM, log~$\tau_c$, log~$\tau_K$, $L_R$, $\dot{E}$, $B_{lc}$, $Q$, W50, W10, and $\alpha$) and report the median p-values from the 1000 sets in Table \ref{tab:ks-pval}.

\begin{table} 
    \centering
    \begin{tabular}{ccc}
    \hline
    \hline
   Parameter & K-S & A-D \\
    \hline
    $P$ & 0.18 & 0.066 \\
    $B_s$ &  0.83 & 0.65 \\
    $\dot{P}$ &  0.89 & 0.74 \\
    DM  & 0.018  & 0.012  \\
    log $\tau_c$  &   0.049 & 0.11 \\
    log $\tau_K$  &  0.52 & 0.57 \\
    $L_{R400}$ &  0.18 & 0.072 \\
    $\dot{E}$ &  0.10 & 0.042 \\
    $B_{lc}$ &   0.20 & 0.022 \\
    $Q$ &  0.10 & 0.023 \\
    W50 &   0.44 &  0.29 \\
    W10 &  0.31  & 0.062 \\
    $\alpha$ &  0.036 &  0.026\\
    \hline
    \end{tabular}
    \caption{P-values resulting from Kolmogorov-Smirnov (K-S) and Anderson-Darling (A-D) two-sample tests comparing the proposed populations of pulsars with $NF<37.5\%$ and pulsars with $NF>37.5\%$ for spin period in seconds ($P$),  magnetic field in Gauss ($B_s$), spindown rate in seconds per second ($\dot{P}$), dispersion measure (DM), characteristic age ($\tau_c$), kinematic age ($\tau_k$), radio luminosity at 400 MHz ($L_{R400}$), spindown luminosity ($\dot{E}$), pulse widths at 50\% (W50) and 10\% (W10), and magnetic inclination angle $\alpha$ in 162 nulling pulsar entries with measured values of $NF (\%)$ taken from \citet{Konar2019}. We run these tests with 1000 sets of corrected NF measurements and report the median p-value. Given the above values, we fail to reject the null hypothesis that the two samples are drawn from the same population.}
    \label{tab:ks-pval}
\end{table}

Given the large p-values resulting from both of these tests, we fail to reject the null hypothesis that nulling pulsars above and below a nulling fraction of $\sim$40\% are drawn from the same population. We do find that the two samples are distinct ($p$ < 0.05) in DM and $\alpha$ and potentially distinct in $\dot{E}$, $B_{lc}$, and $Q$. Although these factors indicate separate populations, the other parameters, such as spin period and pulse width, have large p-values even though they are correlated with nulling fraction (see Section~\ref{sec: correlations}). This indicates that the apparent difference in populations is an artifact of selection bias or a noise limit. We additionally run these two-sample tests on the uncorrected NF measurements. We find that the two uncorrected samples are distinct in DM,  log $\tau_c$,  $L_{R400}$,  $\dot{E}$, $Q$, and $\alpha$, but again not distinct in the more characteristic parameters. We therefore conclude that the two sets of pulsars are drawn from the same population even though they are statistically distinct in some parameters.

\section{An Analysis of Morphological Classifications}
\label{sec: morphology}

We divide our sample into the morphological classifications of pulse profile \citep{Rankin1983}. Using Kolmogorov-Smirnov and Anderson-Darling tests, we investigate whether the populations are statistically distinct in nulling fraction and characteristic age, both of which have been claimed in the literature \citep[e.g., ][]{Hesse1974, Rankin1986}. 

We find that the relation between NF and morphological class is equivocal at best. Pulsars with double profiles (class ``D'') are a statistically distinct population in NF when compared to all other classes, with p-values $<0.05$, except multiple profiles (class ``M''), but no other classes are statistically distinct from each other.

\begin{figure} 
    \centering
    \includegraphics[width=0.45\textwidth]{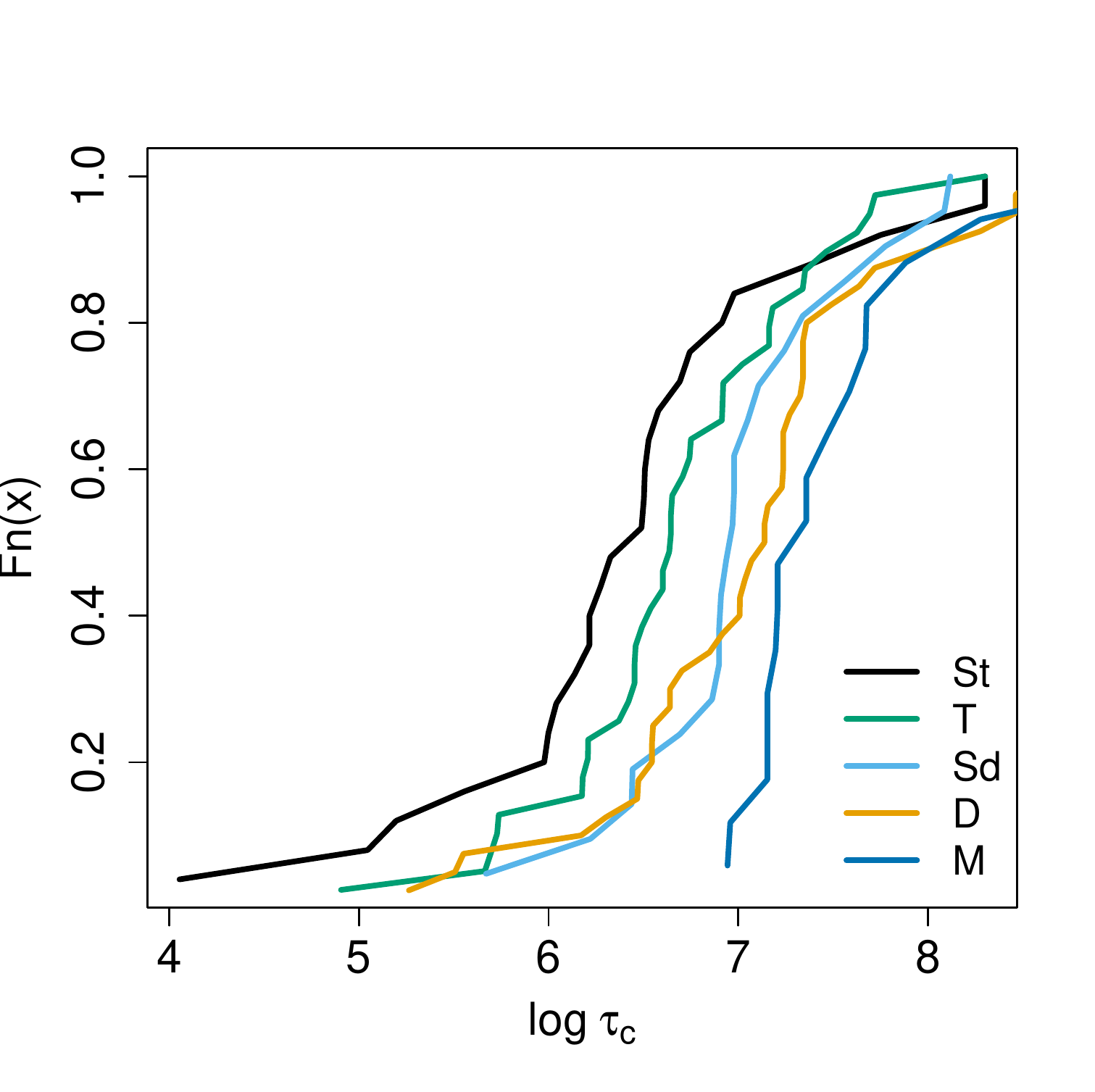}
    \caption{Cumulative Distribution Functions of characteristic age for the five morphological classifications S$_t$, T, S$_d$, D, and M. We find classes to be distinct from other classifications, but cannot prove that neighboring classifications (e.g. S$_t$ and T) are distinct from one another via K-S or A-D tests.}
    \label{fig:tauc_cdf}
\end{figure}

The relation between characteristic age and morphological class is extant in our sample, but subtle. We find that S$_t$ nulling pulsars are the youngest morphological class by median and mean, followed by $T$, then S$_d$, then D, then M, which is consistent with the age-morphology findings in \citet{Rankin1986}. However, in characteristic age, we cannot prove that neighboring classifications are distinct (e.g., S$_t$ and T), but any given class is statistically distinct from all non-neighboring classes (e.g., S$_t$ is distinct from S$_d$, D, and M). The gradient of age across profile classes can be seen in Figure~\ref{fig:tauc_cdf}, but neighboring CDFs are too similar to be distinguished by the K-S and A-D tests.

%%%%%%%%%%%%%%%%%%%%%%%%%%%%%%%%%%%%%%%%%%%%%%%%%%%%%%%%%%%%%%%%%%%%%%%%%%%%%%%%%%%%%%%%%%%%%%%%%%%%%%%%%%%%%%%%%%%%%%%%%%%%%
%%%%%%%%%%%%%%%%%%%%%%%%%%%%%%%%%%%%%%%%%%%%%%            CORRELATIONS            %%%%%%%%%%%%%%%%%%%%%%%%%%%%%%%%%%%%%%%%%%%%%
%%%%%%%%%%%%%%%%%%%%%%%%%%%%%%%%%%%%%%%%%%%%%%%%%%%%%%%%%%%%%%%%%%%%%%%%%%%%%%%%%%%%%%%%%%%%%%%%%%%%%%%%%%%%%%%%%%%%%%%%%%%%%

\section{Correlations Between Pulsar Parameters Within the Nulling Population}
\label{sec: correlations}

\subsection{Searching for Correlations with Null Fraction}
\label{ssec: null_correlations}

We calculate the Pearson-R correlation value and the Maximal Information Coefficient (MIC) between nulling fraction and each of the fifteen parameters described in Section \ref{sec:data}. 

We determine the significance of the MIC for a parameter $x$ by permuting the NF and $x$ values within 1,000 bootstrapped sub-samples of the dataset. We then calculate the p-value as the percentage of permuted outcomes that meet or exceed the median MIC value of the unscrambled and uncertainty-adjusted (corrected NF) data (see Section~\ref{ssec: upperlims_uncertainties}). We consider correlations with p-values less than 0.05 as statistically significant.

In the past, only linear correlations have been investigated for nulling pulsars, but in this paper we are also able to investigate non-linear correlations by using the MIC (Section \ref{ssec: r_mic}). The difference between the two correlation measures (MIC - $R^2$) describes the non-linearity of a correlation. 

We find no correlations shown by Pearson-R between NF and any other variable. However, MIC reveals weak non-linear correlations between NF and: $P$ (MIC = 0.328, p-value = 0.009), $\Dot{E}$ (MIC = 0.334, p-value = 0.005), $B_{lc}$ (MIC = 0.326, p-value = 0.007), $Q$ (MIC = 0.306, p-value = 0.012), and W10 (MIC = 0.337, p-value = 0.003). 

\begin{figure} 
    \centering
    \includegraphics[width=0.45\textwidth]{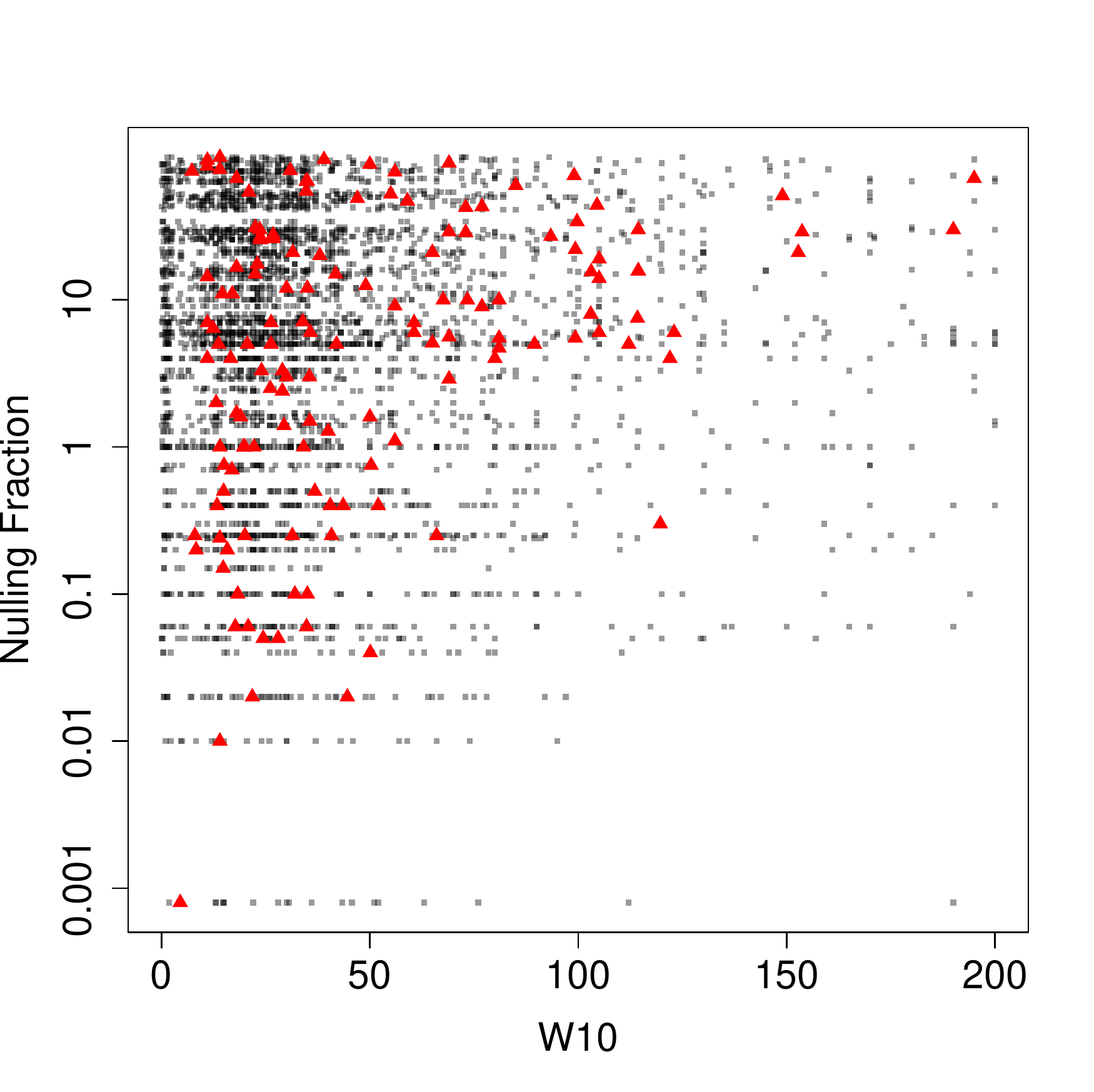}
    \caption{Nulling fraction versus the pulse width (in milliseconds) for 162 nulling pulsar measurements in red triangles. Nulling fractions that have only been measured as upper limits are displayed as this upper limit. For context, we plot the periods of the non-nulling populations against randomly assigned nulling fractions from the nulling population in semi-transparent grey squares to show the expected density with no correlation. Given that the randomly sampled non-nulling pulsar population shows a similar distribution to the nulling pulsar distribution, the correlation is most likely due to selection effects. However, the paucity of points in the lower right quadrant is different between the two distributions and could be suggestive of a scientifically interesting trend.}
    \label{fig:width_nf}
\end{figure}

\begin{figure} 
    \centering
    \includegraphics[width=0.45\textwidth]{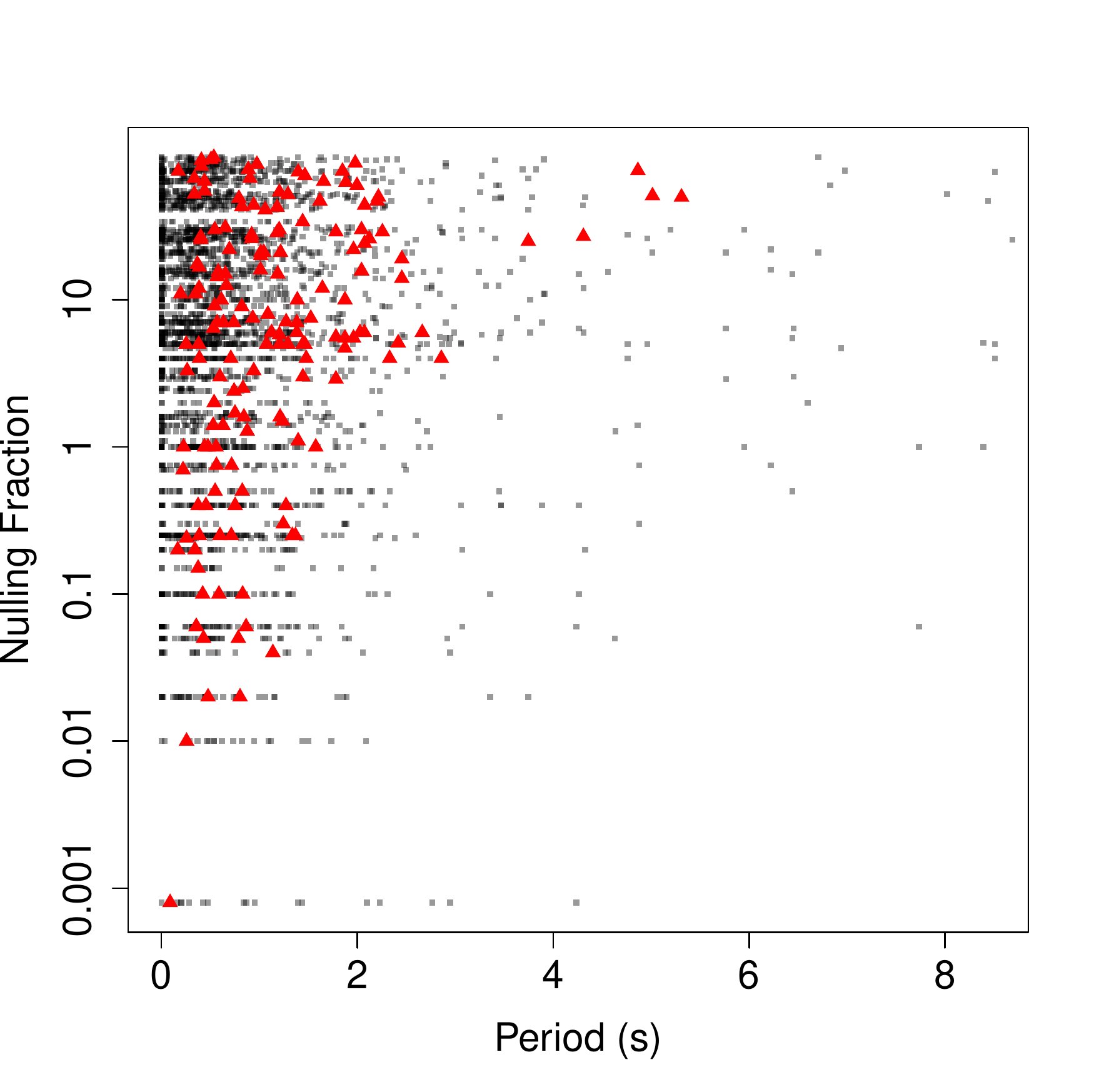}
    \caption{Nulling fraction versus the pulse period for 162 nulling pulsar measurements in red triangles. Nulling fractions that have only been measured as upper limits are displayed as this upper limit. For context, we plot the periods of the non-nulling populations against randomly assigned nulling fractions from the nulling population in semi-transparent grey squares to show the expected density with no correlation. Since the randomly sampled and scrambled non-nulling pulsar population shows a similar distribution to the nulling pulsar distribution, the correlation is most likely due to selection effects.}
    \label{fig:period_nf}
\end{figure}

The correlations with $\Dot{E}$, $B_{lc}$, and $Q$ are not particularly interesting because all three parameters probably take their correlation from their heavy $P$-dependence. However, the weak non-linear trends with $P$ and pulse width align with reports in the previous literature \citep{biggs1992analysis,Li1995} and could point to fundamental physical nulling behaviour. We show the median NF values (for those with upper limits) in Figures \ref{fig:width_nf} and \ref{fig:period_nf}. The measured trends might be driven by selection effects, as illustrated by the formation of very similar distributions by the nulling pulsar population and randomly scrambled samples from the non-nulling pulsar population in Figures \ref{fig:width_nf} and \ref{fig:period_nf}. We therefore conclude that, while these correlations are statistically significant, they are not necessarily scientifically significant and caution against drawing conclusions from them until we have more data. In the meantime, future theories of pulsar nulling should account for these trends with period and pulse width. 

To determine whether our method for accounting for the uncertainties was affected by the choice of distribution, we also perform the same test using a uniform distribution for the uncertainty adjustment instead of a log uniform distribution for NFs reported as an upper limit. We find that two additional correlations, with $L_{R400}$ and $W50$, are only significant in the uniform and log uniform distributions, respectively. Because these correlations are dependent on the uncertainties, and there is no clear scientific basis to choose one distribution over the other, we do not report these parameters as correlated as to NF.

We find numerous correlations between NF and the other parameters \textit{within} morphological classes. However, as the sample sizes are extremely small (13--36 pulsars), we note only that all classes uniformly show statistically significant correlations with the parameters found for the entire population: $P$, $\Dot{E}$, $Q$ , W10 and W50, with the exception of $B_{lc}$ which is correlated with NF in all classes except D. We also recover two potentially interesting correlations not seen in the overall population: $B_s$ and $\dot{P}$ are non-linearly correlated with NF in all classes except S$_t$. However, because $B_s$ is derived from $P$ and $\dot{P}$, this correlation is probably just reflective of a trend with $\dot{P}$.

\subsection{Searching for Correlations between Other Parameters}
\label{ssec: other_correlations}

We calculate the Pearson-R correlation value and the Maximal Information Coefficient (MIC) for every combination of the fifteen parameters described in Section \ref{sec:data}, with the exception of morphology. 

Many correlations between pulsar parameters within the nulling pulsar dataset are expected because the quantities are derived from one another. For example, we expect to see (and do see) a correlation between period $P$ and characteristic age $\tau_c$. This class of derived correlations accounts for the majority of correlations that we find, and while scientifically unoriginal, is still valuable to confirm our methods.

We find a weak but significant non-linear correlation between dispersion measure DM and surface magnetic field $B_s$ (MIC = 0.372, p-value = $4\times10^{-4}$), however we do not see this correlation in the overall non-nulling population. There is a paucity of pulsars with low surface magnetic fields at large DMs, suggesting that the correlation we find is due to a detection bias against distant, low energy pulsars. The nulling population is most likely more sensitive to this bias because nulling pulsars must be observed for a longer span of time to reach the same signal-to-noise ratio as non-nulling pulsars during pulsar discovery surveys.

There is also a weak but significant non-linear correlation between spindown rate $\dot{P}$ and width of the pulse W10 (MIC = 0.318, p-value = 0.006), a correlation that we also see in the non-nulling population. This relationship has been previously reported within alternating spin-down states in individual pulsars \citep[e.g., ][]{Lyne2010, Perera2016}.

%%%%%%%%%%%%%%%%%%%%%%%%%%%%%%%%%%%%%%%%%%%%%%%%%%%%%%%%%%%%%%%%%%%%%%%%%%%%%%%%%%%%%%%%%
%%%%%%%%%           NULLING VS NON-NULLING         %%%%%%%%%%%%%%%%%%%%%%%%%%%%%%%%%%%%%%%%%%%%%%%%%%%%%%%%%%%%%%%%%%%%%%%%%%%%%%%%%%%%%%%%%%

\section{The Nulling vs. Non-Nulling Population}
\label{sec: nullvsnonnull}

\begin{figure*} 
    \centering
    \includegraphics[width=\textwidth]{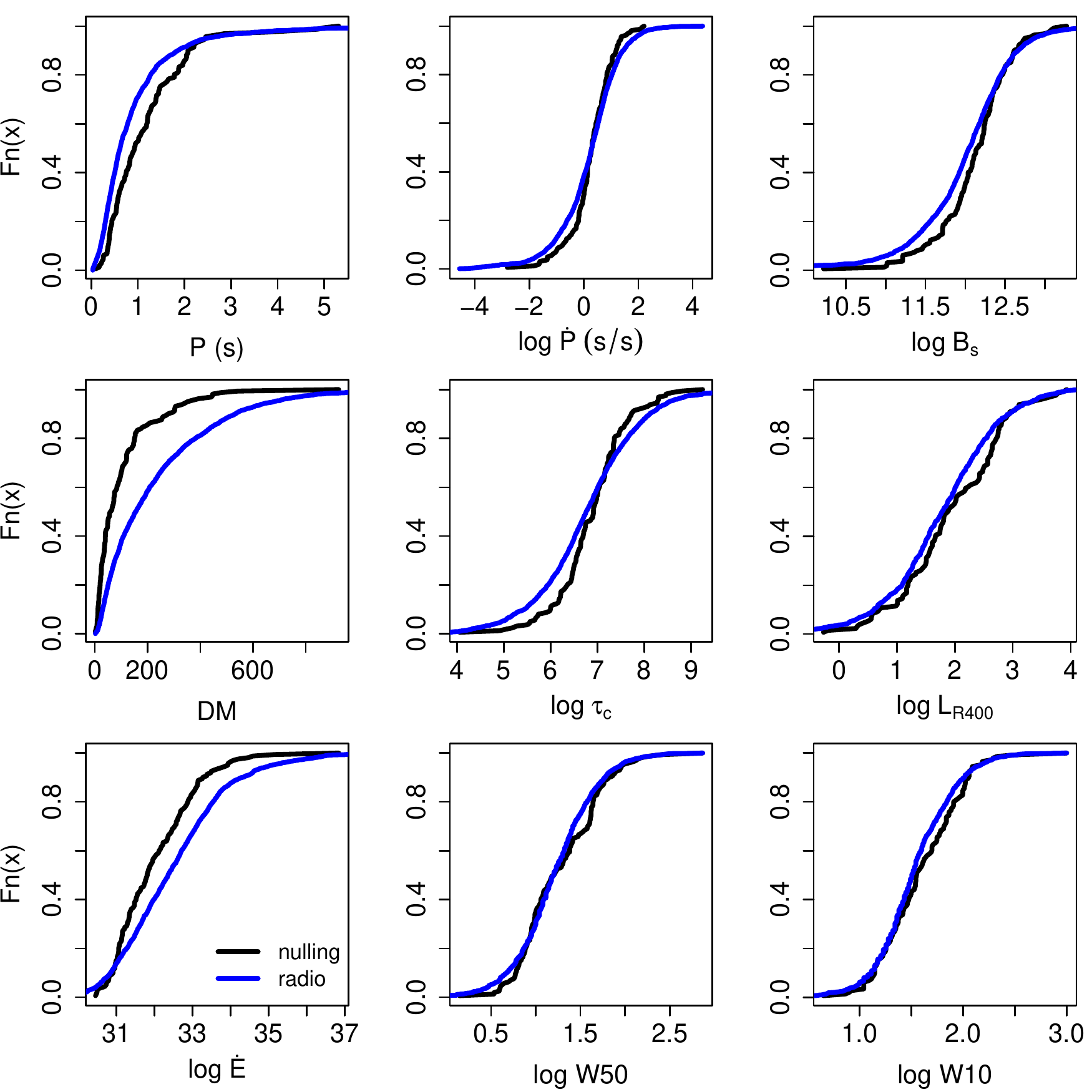}
    \caption{Cumulative distribution functions comparing nulling pulsars in black to normal, radio pulsars in blue. Here,  $P$ is spin period, $\dot{P}$ is spin-down rate, $B_s$ is the surface magnetic field strength, DM is the dispersion measure,    $\tau_c$ is characteristic age,  $L_R400$ is the radio luminosity at 400 MHz, $\dot{E}$ is the spin-down luminosity, $W50$ is the pulse width at 50\% brightness,  and $W10$ is the pulse width at 10\% brightness. We conclude that these two populations are indeed two populations and not samples from a singular population.}
    \label{fig:null_cdfs}
\end{figure*}

To verify that nulling and normal, radio pulsars are two distinct populations, we perform Kolmogorov–Smirnov (K-S) and Anderson-Darling (A-D) two-sample tests, comparing 162 nulling pulsar entries (see Section~\ref{sec:data}) with 2307 radio pulsars in nine parameters that are available via ATNF \citep{Manchester_2005}. We present the resulting p-values from both tests in Table~\ref{tab:nullvsnonnull}. We find that p-values from all parameters except $W10$ and $W50$ are less than $\alpha = 0.05$ and are distinct from one another. We therefore  reject the null hypothesis and confirm that nulling pulsars are not a subset of normal, radio pulsars but are instead a separate population.  We compare the cumulative distribution functions for each parameter between the two populations in Figure~\ref{fig:null_cdfs}.

\begin{table*} 
\centering
\begin{tabular*}{0.45\linewidth}{ccccc}
\hline
\hline
Parameter & No. Nulling   & No. Radio   & K-S & A-D \\
\hline
    $P$         & 162  & 2307  & 1.313e-05   & 4.196e-09 \\
    $\dot{P}$   & 162  & 1958  & 0.03879   & 0.02330 \\
    $B_s$       & 162  & 1958  & 0.007411   & 0.005287 \\
    DM          & 162  & 2306  & 4.408e-14  & 7.058e-23 \\
    $\tau_c$    & 162  & 1958  & 0.002137   & 0.001875 \\
    $L_R$       & 131  & 700   & 0.0293   & 0.04452 \\
    $\dot{E}$   & 161  & 1958  & 0.0002043   & 1.714e-06 \\
    $W50$       & 158  & 1971  & 0.0363   & 0.09156 \\
    $W10$       & 145  & 1154  & 0.1802   & 0.08264 \\ 
\hline
\end{tabular*}
\caption{Resulting p-values from Kolmogorov–Smirnov (K-S) and Anderson-Darling (A-D) two-sample tests. Here,  $P$ is spin period, $\dot{P}$ is spin-down rate, $B_s$ is the surface magnetic field strength, DM is the dispersion measure,    $\tau_c$ is characteristic age,  $L_R$ is the radio luminosity at 400 MHz, $\dot{E}$ is the spin-down luminosity, $W50$ is the pulse width at 50\% peak brightness,  and $W10$ is the pulse width at 10\% brightness. Given our small p-values, we reject the null hypothesis that these two samples are from the same population, even though we cannot claim they are distinct in $W10$ or $W50$. We instead conclude that nulling pulsars are not a subset of radio pulsars and are a distinct and separate population. See Figure \ref{fig:null_cdfs} for the cumulative distribution functions of the two populations in all nine parameters.}
\label{tab:nullvsnonnull}
\end{table*}

%%%%%%%%%%%%%%%%%%%%%%%%%%%%%%%%%%%%%%%%%%%%%%%%%%%%%%%%%%%%%%%%%%%%%%%%%%%%%%%%%%%%%%%%%%%%%%%%%%%%%%%%%%%%%%%%%%%%%%%%%%%%%
%%%%%%%%%%%%%%%%%%%%%%%%%%%%%%%%%%%%%%%%%%%%%%            DISCUSSION            %%%%%%%%%%%%%%%%%%%%%%%%%%%%%%%%%%%%%%%%%%%%%
%%%%%%%%%%%%%%%%%%%%%%%%%%%%%%%%%%%%%%%%%%%%%%%%%%%%%%%%%%%%%%%%%%%%%%%%%%%%%%%%%%%%%%%%%%%%%%%%%%%%%%%%%%%%%%%%%%%%%%%%%%%%%

\section{Discussion}\label{sec:discussion}

To put this work in context, we compile the correlations that have been claimed or suggested by all nulling pulsar statistical studies, including this one, in Table~\ref{tab:pulsar_studies}. Table~\ref{tab:pulsar_studies} highlights the lack of a consensus in the literature about correlations between nulling fraction (NF) and intrinsic pulsar parameters. We are the only study to have searched for correlations in every previously suggested parameter (see Section~\ref{sec:data} for a more thorough description of the included parameters). Only recently have enough nulling pulsars been discovered to draw statistically significant conclusions about the population --- this can be seen in the chronological ordering of Table~\ref{tab:pulsar_studies}.

\begin{table*} 
\centering
\begin{tabular}{l|c|c|c|c|c|c|c|c|c|c|c|c|c|c}
\hline
\hline
Study & Sample Size & $P$   & $\Dot{P}$ & $B_{s}$ & $\tau_C$ & $\alpha$ & $L_R$ & $\Dot{E}$ & $B_{lc}$ & $Q$ & $M$ & $\tau_k$ & $W_{P}$ & DM \\
\hline
\citet{Hesse1974} & 15 &  &  & \cellcolor{blue!25} \checkmark &  &  &  &  &  &  & \cellcolor{blue!25} \checkmark &  &  & \\
\hline
\citet{ritchings1976pulsar} & 32 & \cellcolor{blue!25} \checkmark & X & & \cellcolor{blue!25} \checkmark & & & & \cellcolor{blue!25} \checkmark & & & & & \\
\hline
\citet{Zhen-ru1981} & 32 & \cellcolor{blue!25} \checkmark &  & \cellcolor{blue!25} \checkmark &  &  & &  & \cellcolor{blue!25} \checkmark  &  &  &  &  & \\
\hline
\citet{Rankin1986} & 59 & X & X & X & X &  & &  & X &  & \cellcolor{blue!25} \checkmark &  &  & \\
\hline
\citet{biggs1992analysis} & 72 &  \cellcolor{blue!25} \checkmark &  & \cellcolor{blue!25} \checkmark & \cellcolor{blue!25} \checkmark & \cellcolor{blue!25} \checkmark &  \cellcolor{blue!25} \checkmark &  \cellcolor{blue!25} \checkmark &  \cellcolor{blue!25} \checkmark &  \cellcolor{blue!25} \checkmark &  \cellcolor{blue!25} \checkmark &  &  & \\
\hline
\citet{Li1995} & 72 &  &  &  &  &  &  &  &  &  & &  & \cellcolor{blue!25} \checkmark & \\
\hline
\citet{Lazaridis2006} & 19 &  &  &  & \cellcolor{blue!25} \checkmark &  &  &  &  &  &  &  &  & \\
\hline
\citet{wang2007pulsar} & 64\textdagger & X &  &  & \cellcolor{blue!25} \checkmark &  &  &  & &  & X &  &  & \\
\hline
\citet{cordes2008rocking} & 69\textdagger &  &  &  &  & \cellcolor{blue!25} \checkmark &  &  & &  &  &  &  & \\
\hline
\citet{Konar2019} & 204* &  & X & X &  &  &  &  &  &  &  &  &  & \\
\hline
This Work & 162 & \cellcolor{blue!25} \checkmark & X & X & X & X & X & \cellcolor{blue!25} \checkmark & \cellcolor{blue!25} \checkmark & \cellcolor{blue!25} \checkmark & X & X & \cellcolor{blue!25} \checkmark & X\\
\hline
\end{tabular}
\caption{Pulsar parameters studied for correlation with nulling fraction in the literature, along with the results of this work. $P$ is spin period, $\Dot{P}$ is spin-down rate, $B_s$ is the surface magnetic field strength, $\tau_C$ is characteristic age, $\alpha$ is the inclination angle between the magnetic and spin axes, $L_R$ is the radio luminosity at 400 MHz, $\Dot{E}$ is the spin-down luminosity, $B_{lc}$ is the magnetic field strength at the light cylinder, $Q$ is a dimensionless plasma flow parameter from \citet{Beskin1988}, $M$ is a categorical variable representing the morphological class of the pulse profile (as per \citet{Rankin1983}), $\tau_k$ is the kinetic age of the pulsar, and $W_p$ is the total pulse width. Cells with an "X" indicate that the authors investigated that particular variable in their study and found no correlation with nulling fraction. Shaded cells with a checkmark indicate that a study suggested a correlation between the given variable and nulling fraction. Unshaded cells indicate that a study did not use that particular parameter. We investigated correlations with MIC and $R^2$ and found that the five correlations that we found were non-linear. Morphology is a categorical variable so we used K-S and A-D tests to determine the association between morphology and nulling fraction. \citet{Li1995} used equivalent width for the pulse width parameter, while we saw a correlation with W10 but not W50. \textdagger The authors did not report the number of nulling pulsars, so we estimated the number from provided figures. *This study also included pulsars that were known to null but did not have an estimated nulling fraction. Only 162 values were used in the correlation analyses.}
\label{tab:pulsar_studies}
\end{table*}

In Section \ref{ssec: confirming_two_pops}, we conclude that pulsars with NF below 37.5\% and pulsars with NF above 37.5\% are drawn from the same underlying population. In some variables, however, we find the two populations to be distinct such as in Dispersion Measure (DM). DM is dependent on distance which reveals a potential observational bias: pulsars that null for longer are less likely to appear in e.g. a drift scan survey, so we would expect discovered pulsars with higher nulling fractions to be those that are brighter and/or closer to Earth. To confirm that our results are not being affected by nulling pulsars with multiple NF measurements,\footnote{Our sample contains 141 individual pulsars, but numerous pulsars have different estimates of nulling fraction. We therefore typically treat each individual nulling fraction measurement as a different pulsar}, we re-run our analysis a few times by randomly choosing a singular measurement of NF for each pulsar and by omitting these pulsars entirely. We find this does not change our results.

\begin{figure}
    \centering
    \includegraphics[width=0.47\textwidth]{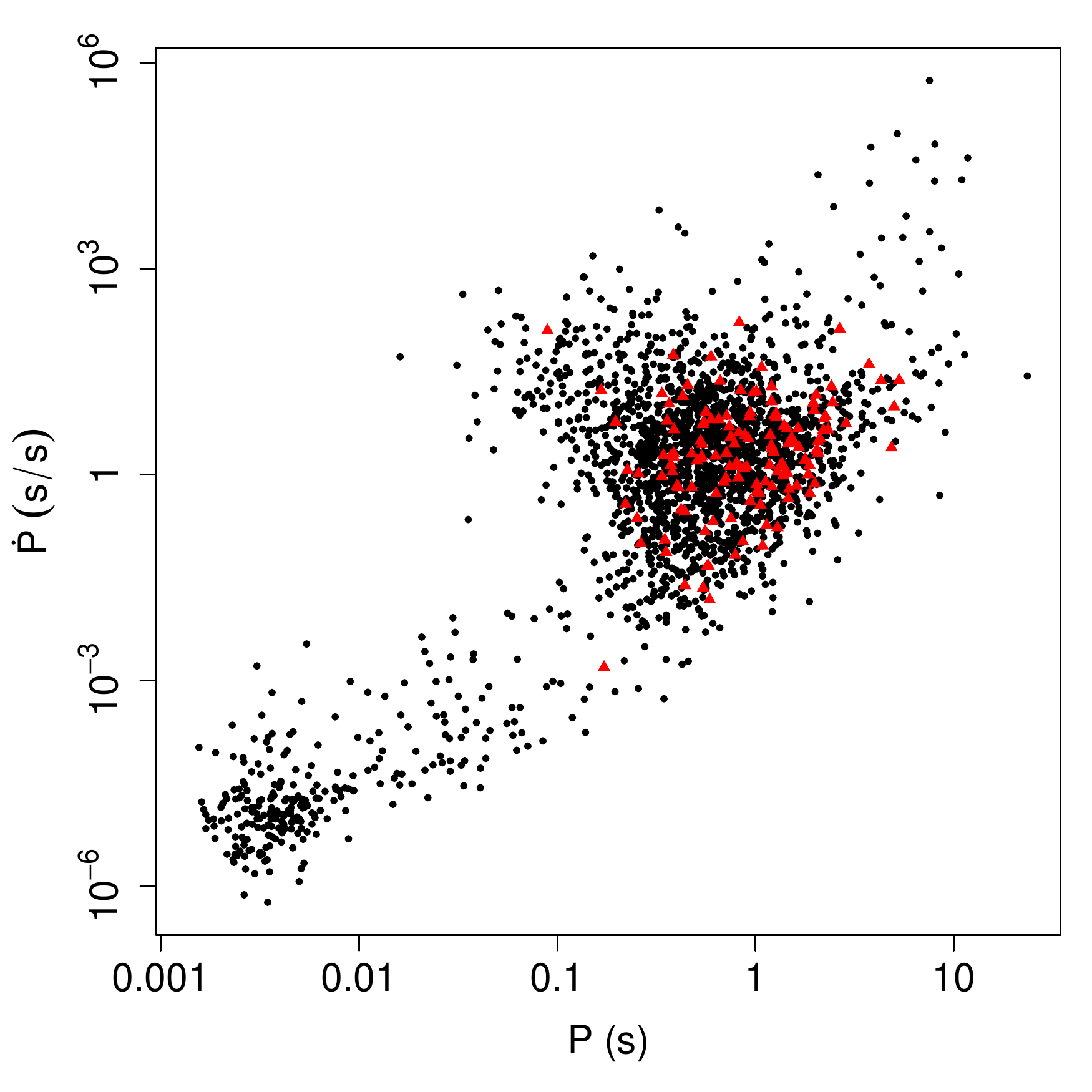}
    \caption{Period vs. Period Derivative plot for non-nulling pulsars (black dots) and nulling pulsars (red triangles). The absence of red triangles in the bottom left of the plot indicates that there are no known millisecond pulsars that null.}
    \label{fig:p_vs_pdot}
\end{figure}

In Figure \ref{fig:p_vs_pdot}, we show a $P$--$\dot{P}$ plot for all of the pulsars in the ATNF database \citep{Manchester_2005}, with nulling pulsars highlighted in red triangles. The ``tail'' in the plot towards the bottom left corner is comprised of the millisecond pulsar population, which has extremely short spin periods and very small $\dot{P}$ values. \citet{rajwade2016probing} suggested that millisecond pulsars might not null at all. The current nulling pulsar population, as shown in Figure \ref{fig:p_vs_pdot} supports this conclusion. 

Given the non-linear correlation between NF and $P$ that we found in this work, the absence of nulling millisecond pulsars points to a fundamental relationship between spin period and nulling behaviour. This motivated the decision to remove both millisecond pulsars and all other non-normal pulsars from the comparison between nulling and non-nulling pulsars in Section~\ref{sec: nullvsnonnull}.

To ensure that the uncertainties resulting from pulse widths measured at different frequencies are not obscuring our results, we perform the following. First, we find that our pulse widths were measured at the following frequencies: roughly one third of the measurements (both W10 and W50) were made at 1400 MHz, one third at 408 MHz, and the rest between those frequencies, with only a single outlier at 5000 MHz (the W10 value for B1641-45). This factor of 3.5 in observing frequency translates to differences in measured pulse widths.  Previous literature shows that, because observing frequency is related to pulse width by a power law, differences in observing frequency of factors $<$10 do not greatly affect pulse widths in most cases \citep[e.g., ][]{DAmico1998parkes,chen2014frequency}. \citet{rankin1983toward2} found that a change of pulse width by a factor of 2 requires multiple orders-of-magnitude changes in observing frequency. To prove that this potential source of error does not affect our results, we add in errors of 60\% \footnote{60\% corresponds to the largest errors from \citet{chen2014frequency}, assuming pulse widths varied by a factor of 10, and $\sim$2.5 greater than we observe in our sample} and perform a bootstrapped analysis. We find that, while the strength of the correlation decreases in some of our trials, our conclusions do not change. We still find a nonlinear, weak, but statistically significant correlation between pulse width and NF.

While we searched for correlations between NF and other parameters within pulse profile morphological classes in Section \ref{ssec: other_correlations}, we were limited by small-number statistics. In order to determine whether nulling behaviour depends on morphological class, a much larger nulling population will be needed. One indirect way to increase the sample of nulling pulsars is to acknowledge the link between mode-changing behaviour and nulling behaviour \citep[e.g., ][]{Timokhin2010}. The fraction of time that a mode-changing pulsar spends in a less-energetic mode can be used as a \textit{de facto} nulling fraction.

Selection effects influence the population of nulling pulsars more than the overall pulsar population, as discussed in \citet{biggs1992analysis}. For example, an absence of emission from a pulsar is easier to detect if the pulsar has a higher flux density (i.e., is nearer to Earth or more luminous). The issue is complicated by an additional selection bias due to lack of sensitivity towards single pulses. For the majority of pulsars, multiple pulses must be integrated in order to recover a pulse profile.  Single pulse nulls can thus easily be missed. Nearer pulsars will have lower DMs, explaining the center-left plot in Figure \ref{fig:null_cdfs}. Thus it is hard to characterize intrinsic differences between nulling and non-nulling pulsar parameters until the selection effects have been characterized. This could perhaps be accomplished via simulated injection-recovery testing.

Nulling fraction is only a single, and admittedly the crudest, measure of pulsar nulling behaviour. \citet{gajjar2017absence} suggested that the lack of a consistent trend of correlation between pulsar parameters and NF was indicative of NF being a poor metric rather than any inherent behavioural cause. Nulling behaviour can also be characterized with nulling length, which has been reported to be correlated with age \citep{Lazaridis2006}. \citet{Yang2014} defines more complex parameters to quantify nulling that relate the lengths of the on and off states, and finds that spin-down rate and spin-down luminosity are correlated with these parameters. Nulling randomness is yet another way to parameterize nulling behaviour \citep{gajjar2017absence}. Polarization parameters and timing noise, as of yet uncompiled for these pulsars, could also be important to characterizing nulling behaviour. Future statistical studies should incorporate more nulling-related parameters once the data are available to do so.

\newpage

\section{Summary and Conclusion}\label{sec:conclusion}

About a tenth of all known pulsars temporarily cease their normal emission in a phenomenon known as ``nulling.'' Here, we investigate this population of 141 pulsars, using a variety of techniques described in Section~\ref{sec:methods}. We first explore whether there is a gap in the nulling fraction (NF) distribution around 40\%, as suggested by \citet{Konar2019}, with cluster analyses. Then we determine if this gap does indeed separate two distinct pulsar populations. Using both Kolmogorov-Smirnov and Anderson-Darling two-sample tests, we then investigate the statistical distinctness of different morphological classes within the overall nulling population. Following the suggestion in past studies that NF is correlated with a variety of pulsar parameters, we test for such correlations between fundamental pulsar parameters (see Section~\ref{sec:data}) and NF using both the Maximal Information Criterion and Pearson's R. We also investigate correlations between all of the parameters in the nulling pulsar population, independent of their behaviour with NF. Lastly, we explore whether nulling pulsars and non-nulling pulsars are distinct populations. Following these analyses, we conclude the following:

\begin{itemize}
    \item There is a local minimum in the nulling fraction of the nulling pulsar population at about 40\%, but we can not identify a natural breakpoint, and this minimum does not divide two statistically distinct populations (Section \ref{sec: populations}).
    \item In terms of nulling fraction, pulsars in morphological class D are statistically distinct from all other classes except M. No other classes are statistically distinct from one another (Section \ref{sec: morphology}).
    \item Morphological class is related to age in the nulling pulsar population, with the classes ordered, youngest to oldest, as S$_t$, T, S$_d$, D, M, which is consistent with \citet{Rankin1986}. This age gradient is statistically distinguishable, but also subtle with large scatter (Section \ref{sec: morphology}).
    \item We find nulling fraction in the nulling pulsar population to be weakly correlated with pulse period $P$ and full pulse width $W10$, as well as with parameters that are derived from $P$ ($\dot{E}$, $B_{lc}$, $Q$), however these trends should be taken with caution because of strong selection effects (Section \ref{ssec: null_correlations}).
    \item We find dispersion measure DM and surface magnetic field $B_s$ to be weakly, non-linearly correlated in the nulling pulsar population but not in the non-nulling population, likely due to selection bias (Section \ref{ssec: other_correlations}).
    \item We find spindown rate $\dot{P}$ and pulse width W10 to be weakly, non-linearly correlated in the nulling pulsar population as well as in the non-nulling population, consistent with previous studies (Section \ref{ssec: other_correlations}).
    \item Nulling and normal, radio pulsar populations are statistically distinct in nine fundamental parameters: period, spindown rate, surface magnetic field, dispersion measure, characteristic age, radio luminosity, spindown luminosity, and 10\% pulse width and 50\% pulse width (Section \ref{sec: nullvsnonnull}). 
\end{itemize}

This list of statistical outcomes might feel disjointed and detached from the physical sample, but each result gives us a clue about either 1) the fundamental physics behind the pulsar nulling phenomenon or 2) the biases that must be accounted for when such analyses are performed. 

Future work must thoroughly investigate the selection effects present in the nulling pulsar sample. This can be achieved via creating synthetic populations and simulating detection rates for pulsars with different nulling fractions, dispersion measures, radio luminosities, etc. for a range of instruments and survey parameters. A study of this magnitude would allow us to disentangle the connection between nulling fraction and inherent pulsar properties from the effect of detection biases. As always, over time, the nulling pulsar sample will become better characterized by the acquisition of polarizations, nulling lengths, null distributions, timing noise, and other data, which will lead to a better understanding of the emission mechanism and its relationship to nulling behaviour.

\section*{Acknowledgements}

We would like to thank the referee for the helpful comments and suggestions that have improved this manuscript. We would like to thank Vishal Gajjar, Alex Wolszczan, and Jason Wright for the helpful discussions. MGM acknowledges that this material is based upon work supported by the National Science Foundation Graduate Research Fellowship Program under Grant No. DGE1255832. Any opinions, findings, and conclusions or recommendations expressed in this material are those of the author and do not necessarily reflect the views of the National Science Foundation. We acknowledge use of the ATNF catalog at \url{http://www.atnf.csiro.au/research/pulsar/psrcat/}. The Center for Exoplanets and Habitable Worlds is supported by the Pennsylvania State University, the Eberly College of Science, and the Pennsylvania Space Grant Consortium.

\section*{Data availability}
The data underlying this article will be shared on reasonable request to the corresponding author.

%%%%%%%%%%%%%%%%%%%%%%%%%%%%%%%%%%%%%%%%%%%%%%%%%%



\bibliographystyle{mnras}
\bibliography{main}

\begin{thebibliography}{}
\makeatletter
\relax
\def\mn@urlcharsother{\let\do\@makeother \do\$\do\&\do\#\do\^\do\_\do\%\do\~}
\def\mn@doi{\begingroup\mn@urlcharsother \@ifnextchar [ {\mn@doi@}
  {\mn@doi@[]}}
\def\mn@doi@[#1]#2{\def\@tempa{#1}\ifx\@tempa\@empty \href
  {http://dx.doi.org/#2} {doi:#2}\else \href {http://dx.doi.org/#2} {#1}\fi
  \endgroup}
\def\mn@eprint#1#2{\mn@eprint@#1:#2::\@nil}
\def\mn@eprint@arXiv#1{\href {http://arxiv.org/abs/#1} {{\tt arXiv:#1}}}
\def\mn@eprint@dblp#1{\href {http://dblp.uni-trier.de/rec/bibtex/#1.xml}
  {dblp:#1}}
\def\mn@eprint@#1:#2:#3:#4\@nil{\def\@tempa {#1}\def\@tempb {#2}\def\@tempc
  {#3}\ifx \@tempc \@empty \let \@tempc \@tempb \let \@tempb \@tempa \fi \ifx
  \@tempb \@empty \def\@tempb {arXiv}\fi \@ifundefined
  {mn@eprint@\@tempb}{\@tempb:\@tempc}{\expandafter \expandafter \csname
  mn@eprint@\@tempb\endcsname \expandafter{\@tempc}}}

\bibitem[\protect\citeauthoryear{Albanese, Riccadonna, Donati  \&
  Franceschi}{Albanese et~al.}{2018}]{albanese2018practical}
Albanese D.,  Riccadonna S.,  Donati C.,   Franceschi P.,  2018, GigaScience,
  7, giy032

\bibitem[\protect\citeauthoryear{Backer}{Backer}{1970}]{backer1970pulsar}
Backer D.,  1970, Nature, 228, 42

\bibitem[\protect\citeauthoryear{Beskin}{Beskin}{2009}]{beskin2009mhd}
Beskin V.~S.,  2009, MHD flows in compact astrophysical objects: accretion,
  winds and jets.
Springer Science \& Business Media

\bibitem[\protect\citeauthoryear{Beskin, Gurevich  \& Istomin}{Beskin
  et~al.}{1984}]{beskin1984spin}
Beskin V.,  Gurevich A.,   Istomin Y.~N.,  1984, Astrophysics and Space
  Science, 102, 301

\bibitem[\protect\citeauthoryear{Beskin, Gurevich  \& Istomin}{Beskin
  et~al.}{1988}]{Beskin1988}
Beskin V.~S.,  Gurevich A.~V.,   Istomin Y.~N.,  1988, \mn@doi [Astrophysics
  and Space Science] {10.1007/BF00637577}, 146, 205

\bibitem[\protect\citeauthoryear{Biggs}{Biggs}{1992}]{biggs1992analysis}
Biggs J.~D.,  1992, The Astrophysical Journal, 394, 574

\bibitem[\protect\citeauthoryear{Bivand}{Bivand}{2019}]{classInt2019}
Bivand R.,  2019, classInt: Choose Univariate Class Intervals.
\url {https://CRAN.R-project.org/package=classInt}

\bibitem[\protect\citeauthoryear{{Chen} \& {Wang}}{{Chen} \&
  {Wang}}{2014}]{chen2014frequency}
{Chen} J.~L.,  {Wang} H.~G.,  2014, \mn@doi [\apjs]
  {10.1088/0067-0049/215/1/11}, \href
  {https://ui.adsabs.harvard.edu/abs/2014ApJS..215...11C} {215, 11}

\bibitem[\protect\citeauthoryear{Cordes \& Shannon}{Cordes \&
  Shannon}{2008}]{cordes2008rocking}
Cordes J.~M.,  Shannon R.,  2008, The Astrophysical Journal, 682, 1152

\bibitem[\protect\citeauthoryear{Coulson}{Coulson}{1987}]{coulson1987matter}
Coulson M.~R.,  1987, Cartographica: The International Journal for Geographic
  Information and Geovisualization, 24, 16

\bibitem[\protect\citeauthoryear{Cox}{Cox}{1972}]{cox1972regression}
Cox D.~R.,  1972, Journal of the Royal Statistical Society: Series B
  (Methodological), 34, 187

\bibitem[\protect\citeauthoryear{{D'Amico}, {Stappers}, {Bailes}, {Martin},
  {Bell}, {Lyne}  \& {Manchester}}{{D'Amico} et~al.}{1998}]{DAmico1998parkes}
{D'Amico} N.,  {Stappers} B.~W.,  {Bailes} M.,  {Martin} C.~E.,  {Bell} J.~F.,
  {Lyne} A.~G.,   {Manchester} R.~N.,  1998, \mn@doi [\mnras]
  {10.1046/j.1365-8711.1998.01397.x}, \href
  {https://ui.adsabs.harvard.edu/abs/1998MNRAS.297...28D} {297, 28}

\bibitem[\protect\citeauthoryear{Fisher}{Fisher}{1958}]{fisher1958grouping}
Fisher W.~D.,  1958, Journal of the American statistical Association, 53, 789

\bibitem[\protect\citeauthoryear{Gajjar}{Gajjar}{2017}]{gajjar2017absence}
Gajjar V.,  2017, arXiv preprint arXiv:1706.05407

\bibitem[\protect\citeauthoryear{Gajjar, Joshi  \& Kramer}{Gajjar
  et~al.}{2012}]{gajjar2012survey}
Gajjar V.,  Joshi B.~C.,   Kramer M.,  2012, Monthly Notices of the Royal
  Astronomical Society, 424, 1197

\bibitem[\protect\citeauthoryear{Gajjar, Joshi  \& Wright}{Gajjar
  et~al.}{2014a}]{GajjarJoshi2014}
Gajjar V.,  Joshi B.~C.,   Wright G.,  2014a, \mn@doi [Monthly Notices of the
  Royal Astronomical Society] {10.1093/mnras/stt2389}, 439, 221

\bibitem[\protect\citeauthoryear{Gajjar, Joshi, Kramer, Karuppusamy  \&
  Smits}{Gajjar et~al.}{2014b}]{Gajjar2014}
Gajjar V.,  Joshi B.~C.,  Kramer M.,  Karuppusamy R.,   Smits R.,  2014b, The
  Astrophysical Journal, 797, 18

\bibitem[\protect\citeauthoryear{Gold}{Gold}{1969}]{gold1969rotating}
Gold T.,  1969, Nature, 221, 25

\bibitem[\protect\citeauthoryear{Goldreich}{Goldreich}{1969}]{goldreich1969physics}
Goldreich P.,  1969, Publications of the Astronomical Society of Australia, 1,
  227

\bibitem[\protect\citeauthoryear{Hankins}{Hankins}{1984}]{hankins1984psr1944+}
Hankins T.,  1984, in Bulletin of the American Astronomical Society. pp
  468--469

\bibitem[\protect\citeauthoryear{Hesse \& Wielebinski}{Hesse \&
  Wielebinski}{1974}]{Hesse1974}
Hesse K.,  Wielebinski R.,  1974, Astronomy and Astrophysics, 31, 409

\bibitem[\protect\citeauthoryear{Huguenin, Manchester  \& Taylor}{Huguenin
  et~al.}{1971}]{huguenin1971properties}
Huguenin G.,  Manchester R.,   Taylor J.,  1971, The Astrophysical Journal,
  169, 97

\bibitem[\protect\citeauthoryear{Jenks}{Jenks}{1977}]{jenks1977optimal}
Jenks G.~F.,  1977, Department of Geographiy, University of Kansas Occasional
  Paper

\bibitem[\protect\citeauthoryear{Jianhua}{Jianhua}{2002}]{xu2002}
Jianhua X.,  2002, Beijing: Higher Education Press, 2002, 37

\bibitem[\protect\citeauthoryear{Jokipii \& Lerche}{Jokipii \&
  Lerche}{1969}]{jokipii1969faraday}
Jokipii J.,  Lerche I.,  1969, The Astrophysical Journal, 157, 1137

\bibitem[\protect\citeauthoryear{Kendall}{Kendall}{1938}]{kendall1938new}
Kendall M.~G.,  1938, Biometrika, 30, 81

\bibitem[\protect\citeauthoryear{Kerr, Hobbs, Shannon, Kiczynski, Hollow  \&
  Johnston}{Kerr et~al.}{2014}]{Kerr2014}
Kerr M.,  Hobbs G.,  Shannon R.~M.,  Kiczynski M.,  Hollow R.,   Johnston S.,
  2014, \mn@doi [Monthly Notices of the Royal Astronomical Society]
  {10.1093/mnras/stu1716}, 445, 320

\bibitem[\protect\citeauthoryear{{Konar} \& {Deka}}{{Konar} \&
  {Deka}}{2019}]{Konar2019}
{Konar} S.,  {Deka} U.,  2019, \mn@doi [Journal of Astrophysics and Astronomy]
  {10.1007/s12036-019-9608-zJApA:
  https://ias.ac.in/article/fulltext/joaa/040/05/0042}, \href
  {https://ui.adsabs.harvard.edu/abs/2019JApA...40...42K} {40, 42}

\bibitem[\protect\citeauthoryear{Kramer, Lyne, O'Brien, Jordan  \&
  Lorimer}{Kramer et~al.}{2006}]{kramer2006periodically}
Kramer M.,  Lyne A.~G.,  O'Brien J.~T.,  Jordan C.~A.,   Lorimer D.~R.,  2006,
  Science, 312, 549

\bibitem[\protect\citeauthoryear{Lazaridis \& Seiradakis}{Lazaridis \&
  Seiradakis}{2006}]{Lazaridis2006}
Lazaridis K.,  Seiradakis J.~H.,  2006, in AIP Conference Proceedings. pp
  309--315, \mn@doi{10.1063/1.2347995}

\bibitem[\protect\citeauthoryear{Leisch}{Leisch}{1999}]{leisch1999bagged}
Leisch F.,  1999, Technical Report~51, Bagged clustering.
Vienna University of Economics and Business Administration

\bibitem[\protect\citeauthoryear{Li \& Wang}{Li \& Wang}{1995}]{Li1995}
Li X.,  Wang Z.,  1995, \mn@doi [Chinese Astronomy and Astrophysics]
  {10.1016/0275-1062(95)00040-Y}, 19, 302

\bibitem[\protect\citeauthoryear{Lyne \& Ashworth}{Lyne \&
  Ashworth}{1982}]{Lyne1982}
Lyne A.~G.,  Ashworth M.,  1982, Monthly Notices of the Royal Astronomical
  Society, 204, 519

\bibitem[\protect\citeauthoryear{Lyne \& Manchester}{Lyne \&
  Manchester}{1988}]{lyne1988shape}
Lyne A.,  Manchester R.,  1988, Monthly Notices of the Royal Astronomical
  Society, 234, 477

\bibitem[\protect\citeauthoryear{Lyne, Ritchings  \& Smith}{Lyne
  et~al.}{1975}]{lyne1975period}
Lyne A.,  Ritchings R.,   Smith F.,  1975, Monthly Notices of the Royal
  Astronomical Society, 171, 579

\bibitem[\protect\citeauthoryear{Lyne, Hobbs, Kramer, Stairs  \& Stappers}{Lyne
  et~al.}{2010}]{Lyne2010}
Lyne A.,  Hobbs G.,  Kramer M.,  Stairs I.,   Stappers B.,  2010, \mn@doi
  [Science] {10.1126/science.1186683}, 329, 408

\bibitem[\protect\citeauthoryear{MacQueen et~al.}{MacQueen
  et~al.}{1967}]{macqueen1967some}
MacQueen J.,  et~al., 1967, Proceedings of the fifth Berkeley symposium on
  mathematical statistics and probability, 1, 281

\bibitem[\protect\citeauthoryear{Malov}{Malov}{1990}]{Malov1990}
Malov I.~F.,  1990, Astronomicheskii Zhurnal, 67, 377

\bibitem[\protect\citeauthoryear{Malov \& Nikitina}{Malov \&
  Nikitina}{2011}]{malov2011angles}
Malov I.,  Nikitina E.,  2011, Astronomy reports, 55, 19

\bibitem[\protect\citeauthoryear{Manchester, Hobbs, Teoh  \& Hobbs}{Manchester
  et~al.}{2005}]{Manchester_2005}
Manchester R.~N.,  Hobbs G.~B.,  Teoh A.,   Hobbs M.,  2005, \mn@doi [The
  Astronomical Journal] {10.1086/428488}, 129, 1993–2006

\bibitem[\protect\citeauthoryear{M{\"u}llner et~al.}{M{\"u}llner
  et~al.}{2013}]{mullner2013fastcluster}
M{\"u}llner D.,  et~al., 2013, Journal of Statistical Software, 53, 1

\bibitem[\protect\citeauthoryear{Naidu, Joshi, Manoharan  \&
  Krishnakumar}{Naidu et~al.}{2018}]{naidu2018detection}
Naidu A.,  Joshi B.~C.,  Manoharan P.,   Krishnakumar M.,  2018, Monthly
  Notices of the Royal Astronomical Society, 475, 2375

\bibitem[\protect\citeauthoryear{Nikitina \& Malov}{Nikitina \&
  Malov}{2017}]{Nikitina2017}
Nikitina E.~B.,  Malov I.~F.,  2017, \mn@doi [Astronomy Reports]
  {10.1134/S1063772917070058}, 61, 591

\bibitem[\protect\citeauthoryear{Ostriker \& Gunn}{Ostriker \&
  Gunn}{1969}]{ostriker1969nature}
Ostriker J.,  Gunn J.,  1969, The Astrophysical Journal, 157, 1395

\bibitem[\protect\citeauthoryear{Peacock}{Peacock}{1983}]{peacock1983two}
Peacock J.,  1983, Monthly Notices of the Royal Astronomical Society, 202, 615

\bibitem[\protect\citeauthoryear{Pearson}{Pearson}{1895}]{pearson1895vii}
Pearson K.,  1895, proceedings of the royal society of London, 58, 240

\bibitem[\protect\citeauthoryear{Perera, Stappers, Weltevrede, Lyne  \&
  Rankin}{Perera et~al.}{2016}]{Perera2016}
Perera B.~B.,  Stappers B.~W.,  Weltevrede P.,  Lyne A.~G.,   Rankin J.~M.,
  2016, \mn@doi [Monthly Notices of the Royal Astronomical Society]
  {10.1093/mnras/stv2403}, 455, 1071

\bibitem[\protect\citeauthoryear{{R Core Team}}{{R Core Team}}{2019}]{R2019}
{R Core Team} 2019, R: A Language and Environment for Statistical Computing.
R Foundation for Statistical Computing, Vienna, Austria, \url
  {https://www.R-project.org/}

\bibitem[\protect\citeauthoryear{Rahadianto, Fariza  \& Hasim}{Rahadianto
  et~al.}{2015}]{rahadianto2015risk}
Rahadianto H.,  Fariza A.,   Hasim J. A.~N.,  2015, in 2015 International
  Conference on Data and Software Engineering (ICoDSE). pp 195--200

\bibitem[\protect\citeauthoryear{Rajwade, Arjunwadkar, Gupta  \& Kumar}{Rajwade
  et~al.}{2016}]{rajwade2016probing}
Rajwade K.,  Arjunwadkar M.,  Gupta Y.,   Kumar U.,  2016, Under Preparation

\bibitem[\protect\citeauthoryear{Rankin}{Rankin}{1983a}]{Rankin1983}
Rankin J.~M.,  1983a, {Toward an Empirical Theory of Pulsar Emission. I.
  Morphological Taxonomy}, \url
  {https://patentimages.storage.googleapis.com/41/b5/ca/69ffeea861af61/US8949899.pdf}

\bibitem[\protect\citeauthoryear{{Rankin}}{{Rankin}}{1983b}]{rankin1983toward2}
{Rankin} J.~M.,  1983b, \mn@doi [\apj] {10.1086/161451}, \href
  {https://ui.adsabs.harvard.edu/abs/1983ApJ...274..359R} {274, 359}

\bibitem[\protect\citeauthoryear{Rankin}{Rankin}{1986}]{Rankin1986}
Rankin J.~M.,  1986, \mn@doi [The Astrophysical Journal] {10.1086/163955}, 301,
  901

\bibitem[\protect\citeauthoryear{Rankin}{Rankin}{1990}]{rankin1990toward}
Rankin J.~M.,  1990, The Astrophysical Journal, 352, 247

\bibitem[\protect\citeauthoryear{Reshef et~al.,}{Reshef
  et~al.}{2011}]{reshef2011detecting}
Reshef D.~N.,  et~al., 2011, science, 334, 1518

\bibitem[\protect\citeauthoryear{Ritchings}{Ritchings}{1976}]{ritchings1976pulsar}
Ritchings R.,  1976, Monthly Notices of the Royal Astronomical Society, 176,
  249

\bibitem[\protect\citeauthoryear{Robitaille et~al.,}{Robitaille
  et~al.}{2013}]{robitaille2013astropy}
Robitaille T.~P.,  et~al., 2013, Astronomy \& Astrophysics, 558, A33

\bibitem[\protect\citeauthoryear{Timokhin}{Timokhin}{2010}]{Timokhin2010}
Timokhin A.~N.,  2010, {A model for nulling and mode changing in pulsars},
  \mn@doi{10.1111/j.1745-3933.2010.00924.x}, \url
  {https://academic.oup.com/mnrasl/article-lookup/doi/10.1111/j.1745-3933.2010.00924.x}

\bibitem[\protect\citeauthoryear{Wang, Manchester  \& Johnston}{Wang
  et~al.}{2007a}]{wang2007pulsar}
Wang N.,  Manchester R.,   Johnston S.,  2007a, Monthly Notices of the Royal
  Astronomical Society, 377, 1383

\bibitem[\protect\citeauthoryear{Wang, Manchester  \& Johnston}{Wang
  et~al.}{2007b}]{Wang2007}
Wang N.,  Manchester R.,   Johnston S.,  2007b, Monthly Notices of the Royal
  Astronomical Society, 377, 1383

\bibitem[\protect\citeauthoryear{Yang, Han  \& Wang}{Yang
  et~al.}{2014}]{Yang2014}
Yang A.,  Han J.,   Wang N.,  2014, \mn@doi [Science China: Physics, Mechanics
  and Astronomy] {10.1007/s11433-014-5534-x}, 57, 1600

\bibitem[\protect\citeauthoryear{Zhen-ru \& Yi}{Zhen-ru \&
  Yi}{1981}]{Zhen-ru1981}
Zhen-ru W.,  Yi C.,  1981, \mn@doi [Symposium - International Astronomical
  Union] {10.1017/s0074180900092937}, 95, 215

\makeatother
\end{thebibliography}

% Don't change these lines
\bsp	% typesetting comment
\label{lastpage}
\end{document}